\documentclass[reprint, aps,superscriptaddress,amsmath,amssym,prl]{revtex4-2}
\usepackage{bm}
\usepackage{float}
\usepackage{graphicx}
\usepackage[usenames,dvipsnames]{color}
\usepackage[normalem]{ulem}
\usepackage[svgnames]{xcolor}
\usepackage{bm}
\usepackage{multirow}
\usepackage{titlesec}
\usepackage[utf8]{inputenc}
\usepackage{booktabs}
\usepackage{array}
\usepackage{url}
\usepackage{tikz} 
\usepackage[colorlinks,linkcolor=blue,citecolor=blue,urlcolor=blue]{hyperref}
\usetikzlibrary{arrows.meta} 
\usetikzlibrary{positioning} 

\begin{document}

	\title{Thermal Hall effect in elemental niobium, a two-gap superconductor}
	
	\author{Jing Zhang}
	\affiliation{Wuhan National High Magnetic Field Center and School of Physics, Huazhong University of Science and Technology,  Wuhan  430074, China}
	
	\author{Xiaokang Li}
	\affiliation{Wuhan National High Magnetic Field Center and School of Physics, Huazhong University of Science and Technology,  Wuhan  430074, China}
	
	\author{Kamran Behnia}
	\affiliation{Laboratoire de Physique et d'\'Etude de Mat\'{e}riaux (CNRS)\\ ESPCI Paris, PSL Research University, 75005 Paris, France }
	
	\author{Zengwei Zhu}
	\email{zengwei.zhu@hust.edu.cn}
	\affiliation{Wuhan National High Magnetic Field Center and School of Physics, Huazhong University of Science and Technology,  Wuhan  430074, China}
	\date{\today}
	
	\begin{abstract}
		
		Niobium holds a pivotal place in superconductivity history: it is not only the elemental superconductor with the highest critical temperature, but also a long-standing candidate for multiband pairing whose evidence has remained controversial. A two-gap scenario was proposed as early as 1959, yet experimental proof stayed elusive. Here, through thermal Hall effect measurements, we unambiguously demonstrate a crossover in transverse thermal transport below $T_{\rm c}$, where the dominant carrier switches from hole-like to electron-like upon cooling. This crossover is a clear hallmark of two distinct superconducting condensates. Fitting our data to a two-gap Bardeen-Rickayzen-Tewordt (BRT) model yields a second energy gap of approximately 0.22 $k_{\rm B}T_{\rm c}$, only 11\% of the dominant gap. This small gap size accounts for the difficulty in resolving the two-gap structure in earlier experiments. Crucially, the temperature at which the electron-like contribution to longitudinal thermal conductivity begins to dominate coincides with the onset of the sign reversal in the thermal Hall coefficient, confirming consistency between the two methods, and providing crucial information on the assignment of the superconducting gaps to their respective hole- or electron-like Fermi surface sheets. These findings not only resolve a longstanding controversy, but also establish that multigap superconductivity is far more common than previously assumed, and demonstrate the thermal Hall effect as a powerful probe for resolving gap multiplicities in superconductors.
		
	\end{abstract}
	
	\maketitle
	
	Soon after the Bardeen-Cooper-Schrieffer (BCS) theory \cite{Bardeen1957} laid the cornerstone for the microscopic mechanism of superconductivity, it was recognized that in the complex electronic structure of transition metals with overlapping $s$ and $d$ bands, the single-gap model was insufficient \cite{Suhl1959,Matthias1955}. As early as 1959, Suhl \textit{et al.} \cite{Suhl1959} explored the consequences of two-band contributions for the superconducting gap structure and conceived a formalism for multiband superconductivity. However, the experimental distinction between the two physical pictures of ``strongly anisotropic single gap'' and ``intrinsic multigap'' has been proven challenging \cite{macvicar1968,HESS1991422,HAHN1998,Boaknin2003,RODRIGO2004306,Fletcher2007,Zehetmayer2010,sanna2022,Alshemi2025}. The predicted features exhibit substantial overlap within the limits of experimental resolution. In many superconductors, the multi-gap/single-gap debate \cite{Ruby2015,khasanov2021,Xu2016,hashimoto2018,sun2018,zhao2024,nag2025,Binning1980,Lin2014,Thieman2018} is still unsettled. Identifying novel discriminating criteria would be extremely helpful.

	The thermal Hall effect refers to the generation of a transverse heat current in response to a perpendicular magnetic field under a longitudinal thermal gradient. Over the past two decades, it has been predominantly employed to probe nodal quasiparticles in unconventional superconductors \cite{Krishana1999YBCO,Cvetkovic2015,altangerel2025,Campillo2026,Zhang2001,Kasahara2005}. Moreover, the thermal Hall conductivity $\kappa_{xy}$ has also been utilized to probe the multiband gap structure of an iron-based superconductor, demonstrating its sensitivity to quasiparticles originating from distinct Fermi surface pockets \cite{checkelsky2012thermal}.


	Niobium (Nb), a multiband transition metal \cite{Mattheiss1970} with a body-centered cubic (bcc) structure, has the highest superconducting critical temperature ($T_{\mathrm{c}}$) among pure elements at ambient pressure and has long been regarded as a multigap superconductor candidate \cite{Suhl1959}. Subsequently, Shen \textit{et al.} observed indications supporting the two-gap model in specific-heat measurements in niobium \cite{shen1965evidence,SUNG1965101}. Later, zero-field thermal conductivity measurements by Carlson \cite{carlson1970anomalous} and tunneling experiments by Hafstrom \cite{hafstrom1970case} also yielded results consistent with a two-gap model. However, opposing views have also persisted. Novotny \cite{novotny1975single} failed to reproduce the low-temperature specific heat anomaly reported by Shen \textit{et al.} \cite{shen1965evidence,SUNG1965101}. Sellers \cite{sellers1973anomalous} ascribed the anomaly in his specific heat data to hydrogen impurities. Anderson's thermal conductivity measurements \cite{Anderson1971} did not confirm Carlson's anomaly, and Almond \cite{Almond1972}, from his ultrasonic experiments, asserted that no evidence for a second gap had been observed. These contradictions highlight the limitations of conventional thermodynamic and transport probes in identifying distinct gap structures. Furthermore, niobium has historically been treated as a benchmark single-band superconductor in the framework of longitudinal magnetothermal transport measurements, and has frequently served as a textbook example of s-wave, nodeless gap behavior \cite{noto1969,Lowell1970,Luo2025,Shakeripour2009}. To date, there is still no consensus on whether niobium is a single-gap or two-gap superconductor.

	
	Here, we present a comprehensive study of the thermal Hall effect in niobium and demonstrate that $\kappa_{xy}$ can effectively probe the Bogoliubov quasiparticles of this conventional superconductor, revealing the multiplicity of its superconducting gaps. We begin by analyzing the normal-state electrical conductivity measured as a function of magnetic field within a two-band model, thereby determining the relative contributions of holes and electrons to the overall transport. Measurements of the magnetothermal conductivity and the thermal Hall effect reveal that the hole and electron contributions exhibit distinct temperature dependences as the system enters the superconducting and mixed states. This behavior strongly suggests that holes and electrons are linked to two distinct superconducting gaps. By fitting the temperature-dependent thermal conductivity with a two-gap Bardeen-Rickayzen-Tewordt (BRT) model \cite{Bardeen1959}, we extract the magnitude of the second energy gap, which is consistent with earlier studies advocating a two-gap superconducting state in niobium. We further explain why no two-gap signature is detectable in the longitudinal magnetothermal conductivity, and show that the temperature at which the sign reversal of $\kappa_{xy}$ occurs coincides with the emergence of the electron-like contribution in $\kappa_{xx}$. This cross-validation of the two thermal probes provides a unified picture of the multiband transport in niobium.

	\begin{figure}
		\centering
		\includegraphics[width=1\linewidth]{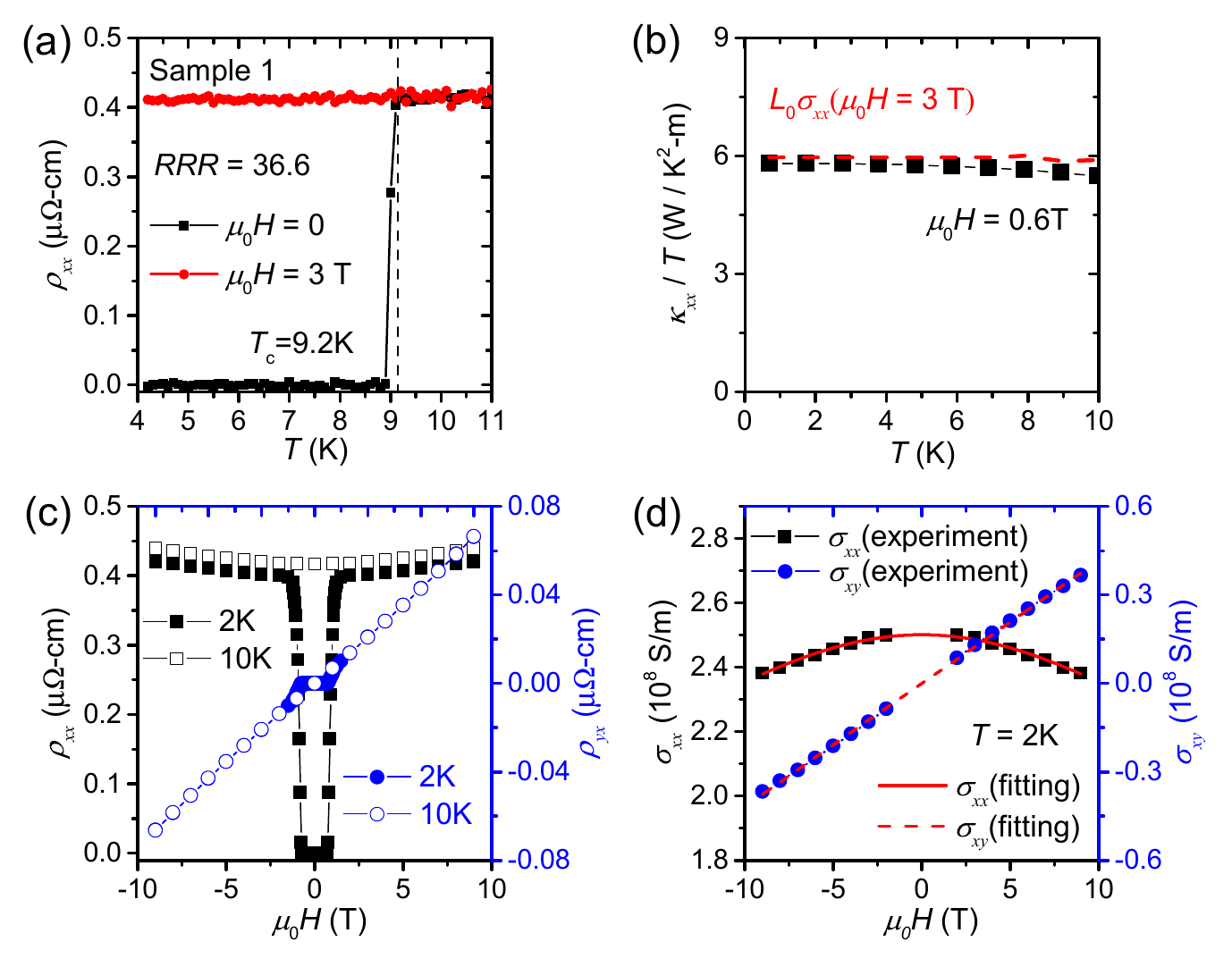} 
		\caption{\textbf{Transport properties in the normal state.} (a) Temperature dependence of resistivity. (b) $\kappa_{xx}/T$ in the normal state, showing quantitative agreement with the Wiedemann–Franz law (red dashed line is $L_0\sigma_{xx}$ at 3 T). (c) Magnetoresistivity $\rho_{xx}(H)$ and Hall resistivity $\rho_{yx}$ at 2 and 10 K. (d) Longitudinal $\sigma_{xx}(H)$ and transverse $\sigma_{xy}(H)$ conductivities at 2 K. The red solid line and dashed line are fits to a two‑band model with the fitting parameters listed in Table~\ref{table1}.}
		\label{fig1}
	\end{figure}
	
	
	\begin{table}
		\centering
		\setlength{\tabcolsep}{9pt}
		\renewcommand{\arraystretch}{1.5}
		\caption{The carrier concentrations and mobilities for holes and electrons at $T$ = 2K. }
		\begin{tabular}{c c c c}
			\toprule
			$n_h$ & $\mu_h$ & $n_e$  & $\mu_e$\\
			$\mathrm{10^{22}\,cm^{-3}}$ & $\mathrm{cm^2/(V\cdot s)}$ & $\mathrm{10^{22}\,cm^{-3}}$ & $\mathrm{cm^2/(V\cdot s)}$ \\
			\midrule
			$6.77$ & $216.5$ & $0.154$ & $621.2$ \\
			\bottomrule
			\label{table1}
		\end{tabular}
	\end{table}
	
	
	We used a commercially available cold-rolled polycrystalline niobium sheet with a thickness of 0.083 mm for our measurements. Fig.~\ref{fig1}(a) shows the temperature dependence of resistivity at zero field and 3 T. The sample exhibits an $RRR = \rho(300~\mathrm{K})/\rho(4.2~\mathrm{K}~\mathrm{at}~3~\mathrm{T})$ of 36.6 and a $T_{\rm c}$ of 9.2 K. Fig.~\ref{fig1}(b) displays the thermal conductivity divided by temperature, $\kappa_{xx}/T$, at 0.6 T, together with $L_0\sigma_{xx}$ at 3 T. Since both $\kappa_{xx}$ and $\sigma_{xx}$ exhibit a very weak magnetic field dependence in the normal state, the excellent agreement between them directly validates the Wiedemann--Franz law. This confirms that heat transport arises predominantly from electronic carriers (holes and electrons), with a nearly negligible phonon contribution, and also validates the reliability of our measurement setup. The electrical conductivities shown in Fig.~\ref{fig1}(d) were extracted by inverting the measured magnetoresistance and Hall resistivity tensors [Fig.~\ref{fig1}(c)], and a two-band model
	$
	\sigma_{xx}(B)=\frac{n_{\rm h} e \mu_{\rm h}}{1+(\mu_{\rm h} B)^2}+\frac{n_{\rm e} e \mu_{\rm e}}{1+(\mu_{\rm e} B)^2},
	\sigma_{xy}(B)=\frac{n_{\rm h} e \mu_{\rm h}^2 B}{1+(\mu_{\rm h} B)^2}-\frac{n_{\rm e} e \mu_{\rm e}^2 B}{1+(\mu_{\rm e} B)^2}
	$
	was used to derive separate carrier concentrations and mobilities for holes and electrons. The values obtained are listed in Table~\ref{table1}. Niobium is found to be uncompensated, with holes as the dominant carriers \cite{Fawcett1967}. The hole density is determined to be \(6.77 \times 10^{22} \, \mathrm{cm}^{-3}\), while the minority electron density is significantly lower at \(0.154 \times 10^{22} \, \mathrm{cm}^{-3}\). This yields a net carrier concentration of \(n_h - n_e = 6.62 \times 10^{22} \, \mathrm{cm}^{-3}\), which is in good agreement with the theoretically predicted value of \(5.56 \times 10^{22} \, \mathrm{cm}^{-3}\) (equivalent to one hole per atom) \cite{Mattheiss1970}. Both holes and electrons also exhibit a relatively low mobility of a few hundred \(\mathrm{cm}^{2} \, \mathrm{V}^{-1} \mathrm{s}^{-1}\), consistent with the polycrystalline nature of the sample.

	\begin{figure}
		\centering
		\includegraphics[width=1\linewidth]{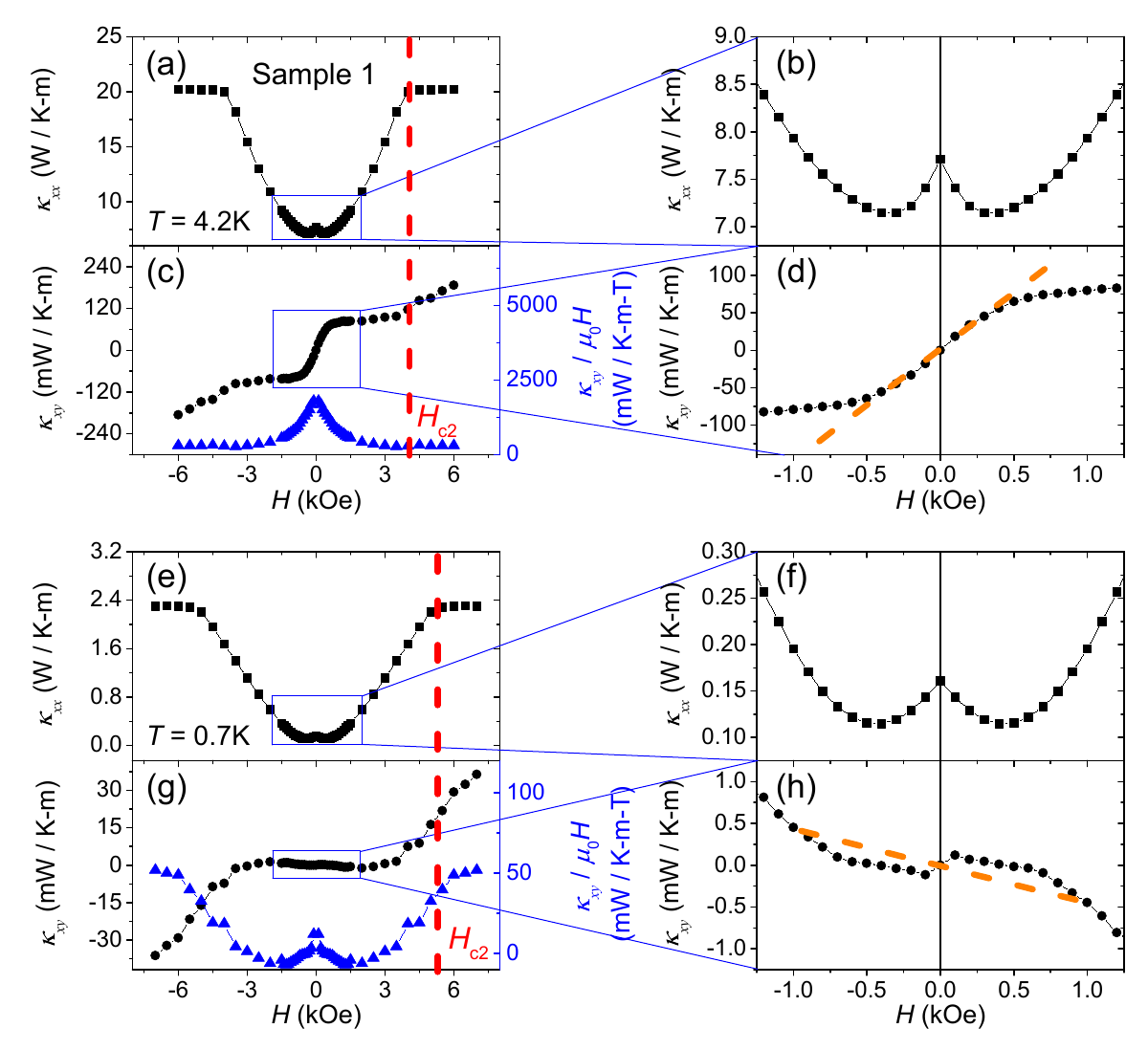} 
		\caption{\textbf{Thermal transport in the superconducting state of niobium at $T = 4.2$ K and $T = 0.7$ K, revealing a sign reversal in the transverse channel from different types of carriers. }(a-d) and (e-h) display $\kappa_{xx}$ and $\kappa_{xy}$ measured at $T = 4.2$ K and $T = 0.7$ K, respectively. The right panels are zoom-ins of the left panels in the low-field region. The orange dashed line indicates the upper critical field $H_{c2}$. The suppression of $\kappa_{xx}$ below 0.5 kOe is attributed to the scattering of quasiparticles and phonons by flux lines. In the low-field region ($<$ 1.25 kOe), the transverse transport is hole-dominated at 4.2 K, whereas it is electron-dominated at 0.7 K, indicated by red eye-guided lines in panel (d) and (h).}
		\label{fig2}
	\end{figure}

	We performed thermal transport measurements using a field-cooling protocol to suppress hysteresis and noise arising from flux pinning. The thermal conducitivity results obtained from the field-scan method and the field-cooling protocol are consistent (see Fig.S6 in Supplementary Material). Fig.~\ref{fig2} displays the field dependence of the longitudinal ($\kappa_{xx}$) and transverse ($\kappa_{xy}$) thermal conductivity measured at 4.2 K and 0.7 K up to 6 kOe, representing two distinct regimes. At 4.2 K (below $T_{\rm c}$) and zero field, quasiparticles dominate the heat transport, while upon cooling to 0.7 K, phonons become the dominant heat carriers (see below). At both temperatures, $\kappa_{xx}$ exhibits a minimum around 0.5 kOe, followed by an increase, and eventually saturates when superconductivity is fully destroyed around 6 kOe. This initial reduction of $\kappa_{xx}$ \cite{kes1975thermal} with magnetic field has been observed in a variety of type-II superconductors \cite{Lowell1970,Chakalskii1978,Behnia1991} and can be attributed to the scattering of the heat carriers (either quasiparticles or phonons) by flux lines. The subsequent increase is driven by the rising quasiparticle density, which provides an additional heat-carrying channel that eventually dominates thermal transport.
	
	At $T = 4.2$ K, $\kappa_{xy}$ increases rapidly in the low-field region ($< 0.4$ kOe), marked by an orange dashed line, with a positive growth rate that even exceeds that of normal state [see Figs.~\ref{fig2}(c) and (d)]. In this low-field region, the quasiparticle density is low, and the mean free path is relatively long. As the field increases further, the quasiparticle density increases sharply around 0.5 kOe, causing the mean free path to drop substantially, and the growth rate of $\kappa_{xy}$ slows down. For fields above $H_{c2}$ [determined by the saturation of $\kappa_{xx}$ in the Fig.~\ref{fig2}(a)], both the heat carrier concentration and the mean free path remain nearly constant, resulting in a linear field dependence of $\kappa_{xy}$.
	
	Remarkably, at 0.7 K, $\kappa_{xy}$ exhibits the opposite sign up to 3 kOe with a much smaller magnitude. The small kink around 0.1 kOe in Fig.~\ref{fig2}(h) may be due to the uncertainty in measuring such small values. However, the negative growth trend is pronounced and evident in Fig.~\ref{fig2}(h), indicated by an orange dashed line. This observation implies that the dominant quasiparticle excitations have changed from holes at 4.2 K to electron-like at 0.7 K in the low-field region. As the field increases beyond 3 kOe, the growth becomes positive, as expected since the dominant hole carriers have been excited by the field and become dominant again. 
	
	The upper critical field $H_{c2}(T)$, determined from the saturation of $\kappa_{xx}$, ie., in Figs.~\ref{fig2}(a, e), follows the empirical parabolic form (see Supplementary Material), indicating the correct reflection of the thermal measurements. 

\begin{figure}
		\centering
		\includegraphics[width=0.75\linewidth]{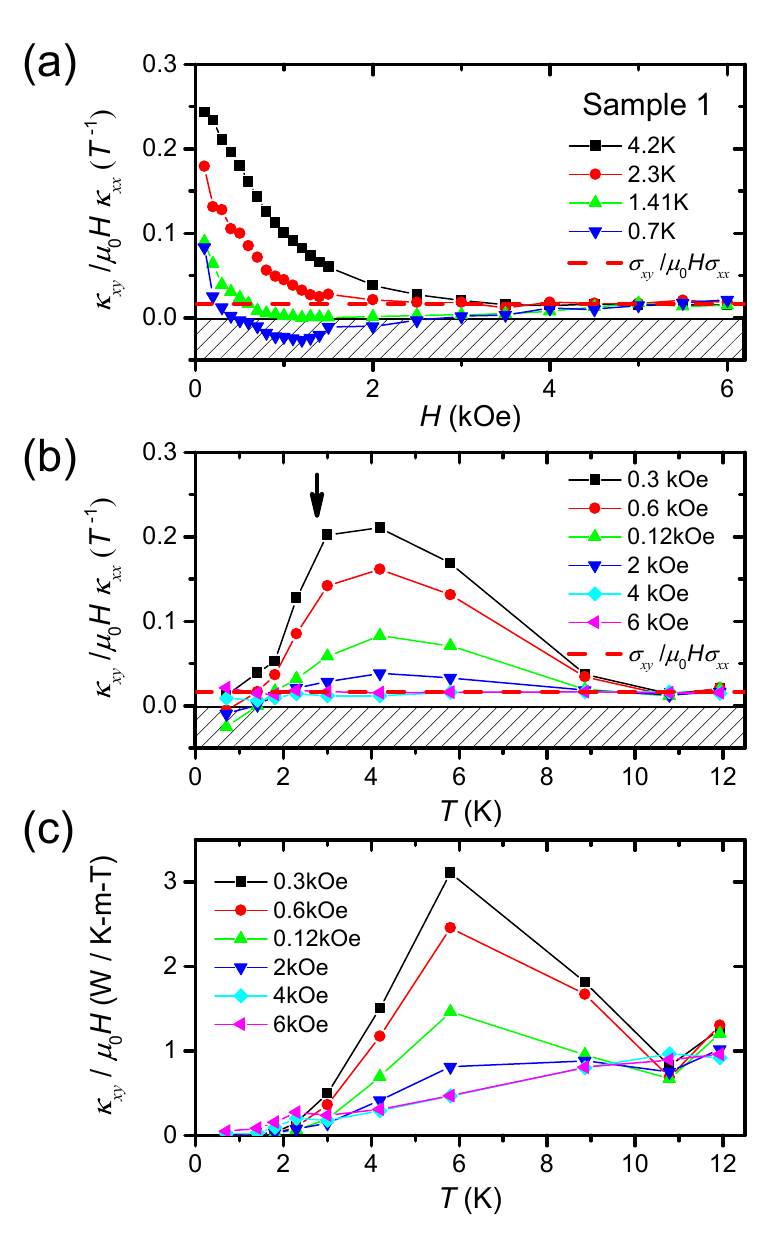} 
		\caption{\textbf{Thermal Hall coefficient versus field (a) and temperature (b) and temperature dependence of the thermal Hall conductivity divided by $\mu_0H$ (c). }The red dashed line represents the normal‑state Wiedemann–Franz law prediction. The shaded region denotes the regime of negative thermal Hall coefficient.  The transverse transport is hole‑type dominated at high fields, whereas at low fields it evolves from hole‑type to electron‑type upon cooling. This suggests that the suppression of electron‑type quasiparticles with decreasing temperature differs from that of hole‑type ones.
		}
		\label{fig3}
	\end{figure}

	Fig.~\ref{fig3} shows the magnetic field and temperature dependences of the thermal Hall coefficient, $\tan\theta_H/\mu_0H = \kappa_{xy}/(\mu_0 H \kappa_{xx})$. In the normal state, the Wiedemann--Franz law is recovered, linking the thermal Hall angle to the ordinary Hall angle. The shaded regions show the negative contribution to $\kappa_{xy}$, arising from the crossover from hole-like to electron-like transport at low fields and low temperatures. Using the single-band quasiclassical transport theory, the Hall angle $\tan\theta_H \sim \omega_c \tau = \frac{eB\tau}{m^*} \sim \frac{eB\ell}{m^*v_F}$, where $e$ is the fundamental charge, $\ell$ is the mean-free-path, $m^*$ is the effective mass, and $v_F$ is the Fermi velocity. Combined with the Wiedemann--Franz law, $\ell\propto\frac{\kappa_{xy}^q}{\mu_0H\kappa_{xx}^q}$, where $q$ denotes hole/electron-type quasiparticles. In the superconducting state, the reduction in quasiparticle scattering when lowering $T$ or $H$ typically leads to a monotonic increase of $\ell$ \cite{altangerel2025}. However, a striking feature of our data is that in the low-field limit $\kappa_{xx}$ decreases upon the entry to the superconducting state but $\kappa_{xy}$ increases[Fig.~\ref{fig3}(c)]. The fact that $\kappa_{xx}$ (see below) decreases (in contrast to cuprates \cite{Krishana1999YBCO,Cvetkovic2015,altangerel2025,Campillo2026,Zhang2001}, organic superconductors \cite{Belin1998,Izawa2001}, and CeCoIn$_5$ \cite{Kasahara2005}) indicates that neither the quasiparticle nor the phonon mean free path increases significantly in the superconducting state. 
	
	Let us focus on the low-field region at 0.3 kOe, where the contribution arises from the intrinsic quasiparticles rather than those excited by the magnetic field. Just below $T_{\rm c}$, the hole quasiparticles (which has a larger gap) are gapped, while the electron quasiparticles are not. The initial increase in $\kappa_{xy}$ is due to the change in this delicate balance between electron and hole contributions. It is worth noting that the extension of the BRT formalism to transverse transport is highly desired but currently unavailable; this scenario (which has not been discussed before) may also contribute to the $\kappa_{xy}$ enhancement reported in other multiband superconductors. 
	
	The nonmonotonic behavior observed in Figs.~\ref{fig3}(a) and (b) confirms the two-band superconducting nature of niobium. Despite the fact that we cannot accurately separate $\kappa_{xy}^i(H)$, it is still clear that the contribution of electrons becomes noticeable below about 3--4~K [indicated by the arrow in Fig.~\ref{fig3}(b) for 0.3 kOe]. The carrier exhibiting a negative transverse thermal response opposite to that of holes becomes dominant only under low-temperature and low-field conditions.
	
	Recently, phonons have been found to contribute to the thermal Hall response \cite{Li2020,grissonnanche2019,grissonnanche2020chiral,boulanger2020,li2023phonon,Jin2025,Boulanger2022}. To exclude phonon contribution in our case, we measured an additional single-crystal Nb sample exhibiting  much larger phonon thermal conductivity (about one order of magnitude larger in the superconducting state). We again observed the sign reversal of the thermal Hall effect upon cooling. However, no enhancement of the thermal Hall coefficient was detected (see Supplementary Material for details), conclusively excluding a phononic origin of the observed thermal Hall response.
	
	\begin{figure*}
		\centering
		\includegraphics[width=0.73\linewidth]{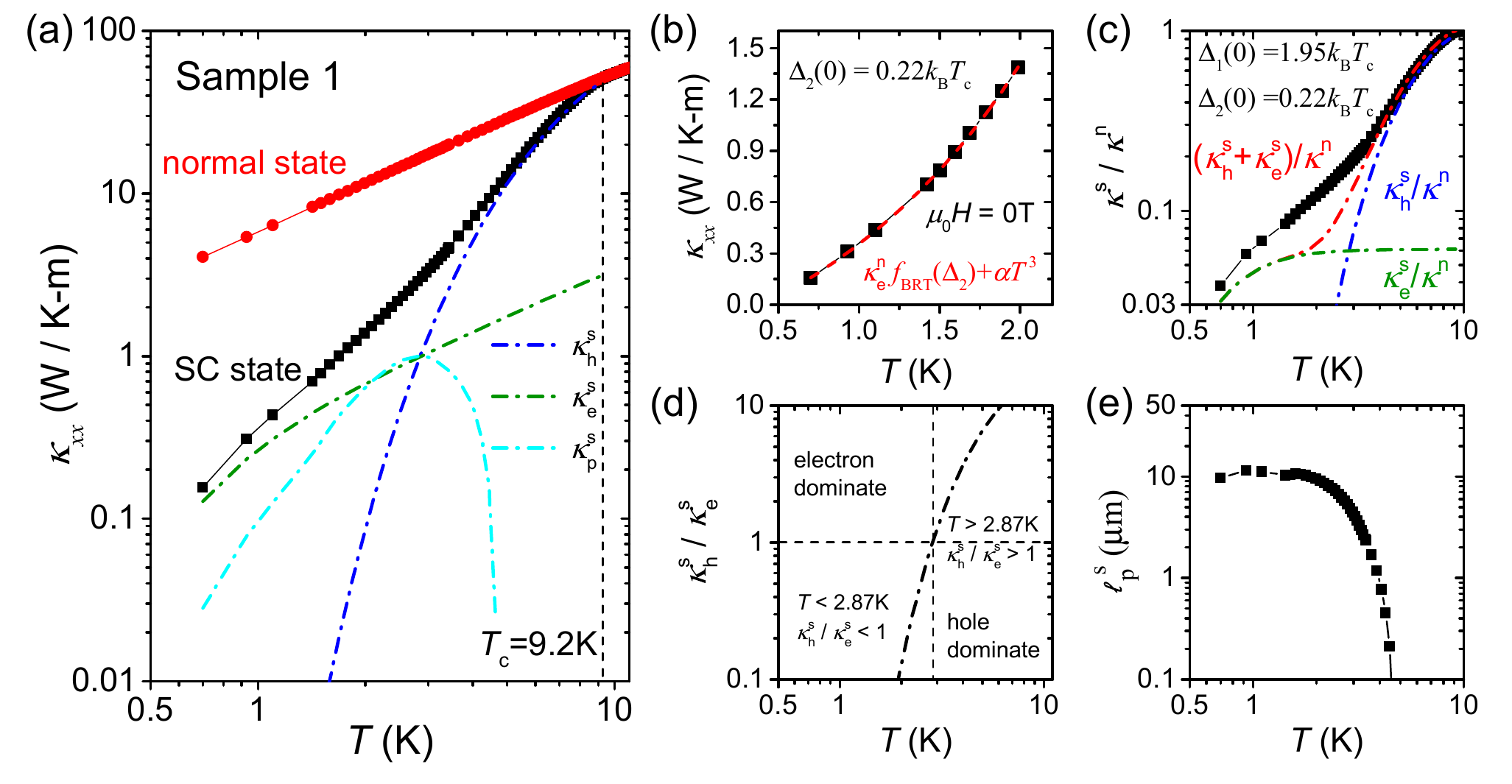} 
		\caption{\textbf{Respective contributions of the carriers at zero field. }(a) Longitudinal thermal conductivity in the normal and superconducting states, with the three resolved superconducting components. $\kappa_h^s$ and $\kappa_e^s$ are from BRT fits; $\kappa_p^s$ is obtained by subtracting these from the measured total. (b) Low‑temperature  fit of $\kappa^s$, considering only $\kappa_e^s$ and $\kappa_p^s$ ($\kappa_h^s$ can be neglected). The $\kappa_p^s$ is assumed to be grain‑boundary scattering limited and proportional to $T^3$. The resulting second gap is $\Delta_2 (0) = 0.22 k_\mathrm{B}T_\mathrm{c}$. (c) Two‑gap BRT fit, based on the assumption that $\kappa_p^s$ is negligible at higher temperatures. The resulting first gap is $\Delta_1(0) = 1.95 k_\mathrm{B}T_\mathrm{c}$. Comparison of our results with previous literature is shown in Table II. (d) Ratio of hole to electron thermal conductivity, indicating the dominant temperature regions. (e) The phonon mean free path. It gradually saturates upon cooling.}
		\label{fig4}
	\end{figure*}
	
	\begin{table*}
		\centering
		\setlength{\tabcolsep}{9pt}
		\renewcommand{\arraystretch}{1.5}
		\caption{Comparison of experimental results on the energy gap in Nb.}
		\begin{tabular}{c c c c}
			\toprule
			Source & Technique & $\Delta_1(0)/k_\mathrm{B}T_\mathrm{c}$  & $\Delta_2(0)/k_\mathrm{B}T_\mathrm{c}$\\
			\midrule
			Sung and Shen \cite{SUNG1965101} & Specific heat & 1.51 & 0.16 \\
			Shen and Senozan $et al.$ \cite{shen1965evidence} & Specific heat & 1.46 & 0.12 \\
			Carlson and Satterthwaite \cite{carlson1970anomalous} & Thermal conductivity & 1.76 & 0.07 \\
			Hafstorm and Macvicar \cite{hafstrom1970case} & Tunneling (Nb/In) & 1.96 & 0.19 \\
			Present work & Thermal Hall effect and thermal conductivity & 1.95(hole)  & 0.22(electron) \\
			\bottomrule
			\label{table2}
		\end{tabular}
	\end{table*}
	
	We now analyze the zero-field longitudinal thermal conductivity to decompose its electronic and phonon contributions and extract the superconducting gap sizes. According to the BRT theory \cite{Bardeen1959}, the ratio of the quasiparticle thermal conductivity in the superconducting state to that in the normal state is given by
	
	\begin{equation}
		\begin{split}
			f_{\mathrm{BRT}}=\frac{\kappa_q^s}{\kappa_q^n} = \frac{1}{f(0)} \Big[ & f(-y) + y \ln\bigl(1+\exp(-y)\bigr) \\
			& + \frac{y^2}{2\bigl(1+\exp(y)\bigr)} \Big]
		\end{split}
		\label{1}
	\end{equation}
	
	Here, $q$ again denotes hole/electron-type quasiparticles, $s$ denotes the superconducting state, and $n$ denotes the normal state. 
	$y = \frac{\Delta(T)}{k_\mathrm{B} T}$, and the Fermi integral 
	$f(-y) = \int_0^\infty \frac{z \, \mathrm{d}z}{1+\exp(z+y)}$. $k_\mathrm{B}$ is the Boltzmann constant, and $\Delta(T)$ is the superconducting gap. According to the interpolation formula for the BCS gap evolution given by F. Gross \cite{gross1986anomalous}, 
	$\Delta(T)=\Delta(0)\tanh{[1.74\sqrt{\frac{T_\mathrm{c}}{T}-1}]}$ and the zero-temperature gap $\Delta(0)$ is the only fit parameter. 
	
	Assuming the energy gap parameter of the hole band is $\Delta_1$ and that of the electron band is $\Delta_2$, the thermal conductivity in the superconducting state is given as
	
	\begin{equation}
		\kappa^s=\kappa_h^s+\kappa_e^s+\kappa_p^s=\kappa_h^n f_{\mathrm{BRT}}(\Delta_1)+\kappa_e^n f_{\mathrm{BRT}}(\Delta_2)+\kappa_p^s
	\end{equation}

	As $T \to T_\mathrm{c}$, the abundant quasiparticle excitations cause the phonon mean free path to become extremely short, such that $\kappa_p^s \to 0$ and $\kappa^s = \kappa_h^s + \kappa_e^s = \kappa_h^n f_{\mathrm{BRT}}(\Delta_1) + \kappa_e^n f_{\mathrm{BRT}}(\Delta_2)$. The normal-state thermal conductivity ratio of holes and electrons $\kappa_e^n / \kappa_h^n$ can be obtained from the Wiedemann-Franz law via $\sigma_{xx}^e / \sigma_{xx}^h = n_e \mu_e / n_h \mu_h$, using the parameters listed in Table~\ref{table1}. At low temperatures ($T \ll T_\mathrm{c}$), where the holes are fully paired ($k_B T_\mathrm{c} \ll \Delta_1$), $\kappa_h^s \to 0$ and phonons are dominantly scattered by grain boundaries, resulting in $\kappa_p^s \propto T^3$. Thus, $\kappa^s = \kappa_e^s + \kappa_p^s = \kappa_e^n f_{\mathrm{BRT}}(\Delta_2) + \alpha T^3$. As shown in Fig.~\ref{fig4}(b), the fitting curve below 2 K agrees well with the experimental data, yielding a small gap of $0.22 k_{\mathrm{B}} T_{\mathrm{c}}$. From a fit to the data above 5 K [Fig.~\ref{fig4}(c)], we obtain a large gap of $1.95 k_{\mathrm{B}} T_{\mathrm{c}}$, which is consistent with the primary superconducting gap values obtained by various methods in the previous literatures \cite{Hirshfeld1962,connolly1962,van1964,Wasim1969,kes1974thermal,mamyia1974thermal,novotny1975single,gladun1977} (Table~\ref{table2}). Importantly, our thermal Hall effect measurements directly reveal the hole- or electron-like nature of these gaps, a crucial piece of information that is not accessible from other methods.
	
	The resulting thermal conductivity components in the superconducting state are shown in Fig.~\ref{fig4}(a), where $\kappa_p^s$ is obtained by subtracting the fitted quasiparticle contribution from the experimental data. Interestingly, the phonon mean free path $\ell_p^s = 3\kappa_p^s / (C_p v_s)$, calculated using specific heat data from Ref.~\cite{shen1965evidence}, gradually increases with decreasing temperature and saturates at approximately 10 $\mu$m [Fig.~\ref{fig4}(e)], which is comparable to the main rolling grooves observed by our SEM analysis (see Supplementary Material).
	
	Fig.~\ref{fig4}(d) shows the ratio of hole to electron quasiparticles, which contributes to thermal conductivity. As the hole quasiparticles associated with the large gap are gapped out, the ratio decreases dramatically with decreasing temperature, and $\kappa_e^s$ begins to dominate below around 2.87 K, $\kappa_h^s/\kappa_e^s$ becomes smaller than 1. Remarkably, the temperature at which the electrons start to prevail is consistent with the temperature at which the thermal Hall coefficient begins to decrease upon cooling at 0.3 kOe in Fig.~\ref{fig3}(b) (marked by the arrow). This crossover confirms that both methods probe the same physical origin—the dominant electron quasiparticle contribution at low temperatures—and demonstrates that the thermal Hall effect is a powerful tool for revealing multiple gaps in superconductors.
	
	Finally, we comment that no evidence of a second energy gap has ever appeared in longitudinal magnetothermal conductivity studies of niobium \cite{Wasim1969,Luo2025,Lowell1970,noto1969}. This is likely due to the small amplitude of the second gap, whose thermal conductivity anomaly signature is confined to very low fields where it is masked by the phonon thermal conductivity (see Supplementary Material). We also note that by carefully analyzing the zero-field data through fitting, the signature of the second gap can still be traced in Anderson’s thermal conductivity data(see Supplementary Material).


	In summary, our thermal Hall effect measurements reveal a crossover of the dominant quasiparticle character from hole-like to electron-like upon cooling, thereby establishing niobium as a two-gap superconductor in which the second gap is associated with the electron band. In contrast to previous determinations based solely on longitudinal thermal conductivity, the inclusion of the thermal Hall effect not only circumvents the complications arising from the phonon background but also provides a direct association between the superconducting gaps and the respective quasiparticle types (holes or electrons). Furthermore, we demonstrate excellent consistency between these two thermal transport probes. Our thermal Hall effect results highlight the need for further theoretical development of the BRT theory to account for transverse transport in multiband superconductors.



	\textit{Acknowledgments---}This work was supported by the National Science Foundation of China (Grant No.12004123, 51861135104 and No.11574097), the Fundamental Research Funds for the Central Universities (Grant no. 2019kfyXMBZ071) and the National Key Research and Development Program of China (Grant No.2022YFA1403503). X. L. acknowledges the China National Postdoctoral Program for Innovative Talents (Grant No.BX20200143) and the China Postdoctoral Science Foundation (Grant No.2020M682386).

	\noindent
	\textit{Data availability---}The data that support the findings of this study are available from the corresponding author upon reasonable request.

%

	
	\clearpage
	
	\begin{center}{\large\bf Supplementary Material for ``Thermal Hall effect in elemental niobium, a two-gap superconductor''}\\
	\end{center}
	
	\renewcommand{\thesection}{S\arabic{section}}
	\renewcommand{\thetable}{S\arabic{table}}
	\renewcommand{\thefigure}{S\arabic{figure}}
	\renewcommand{\theequation}{S\arabic{equation}}
	
	\setcounter{section}{0}
	\setcounter{figure}{0}
	\setcounter{table}{0}
	\setcounter{equation}{0}
	
	\section{Samples and Methods}
	
\begin{figure}[ht]
		\centering
		\includegraphics[width=0.6\linewidth]{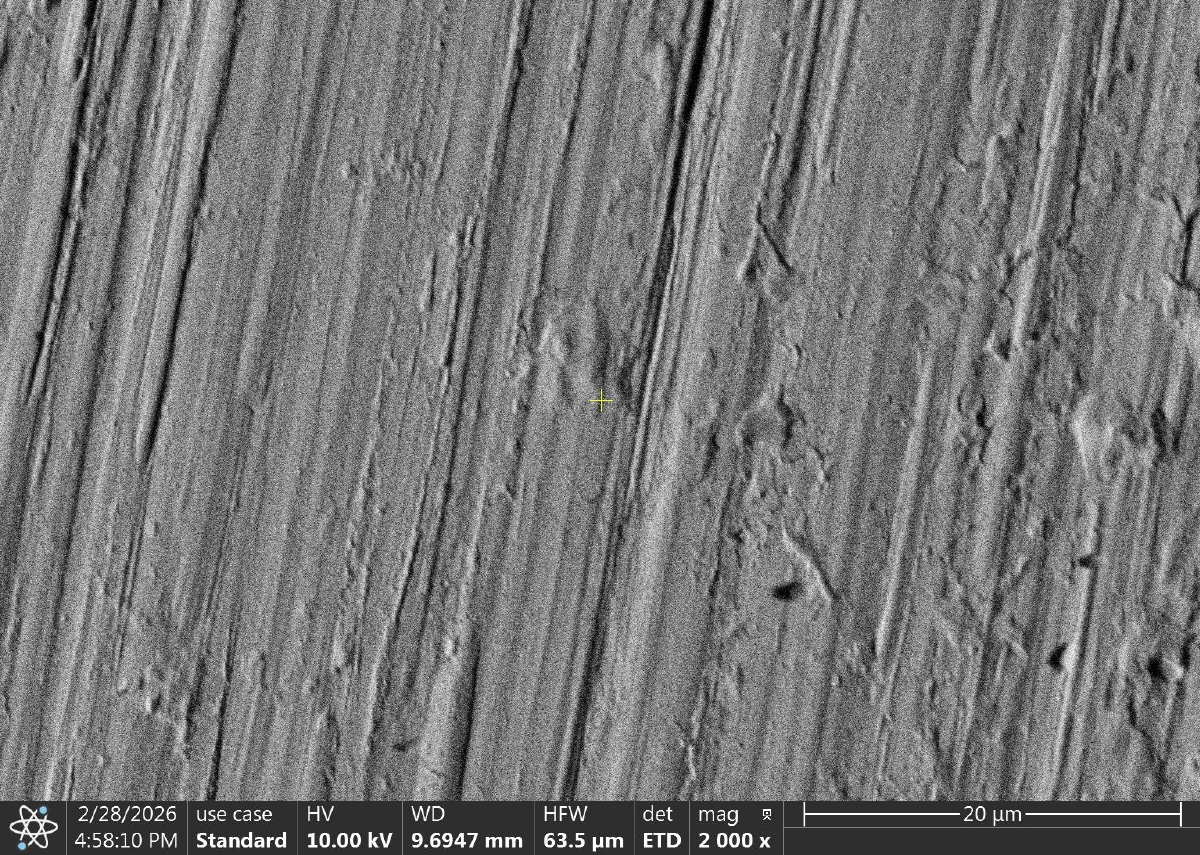} 
		\caption{\textbf{SEM image of the polycrystal sample. }Owing to the cold-rolling process, the main rolling grooves are elongated with a width on the order of 10 $\mu $m.}
		\label{FIG. S1}
	\end{figure}

\begin{figure}[ht]
		\centering
		\includegraphics[width=0.5\linewidth]{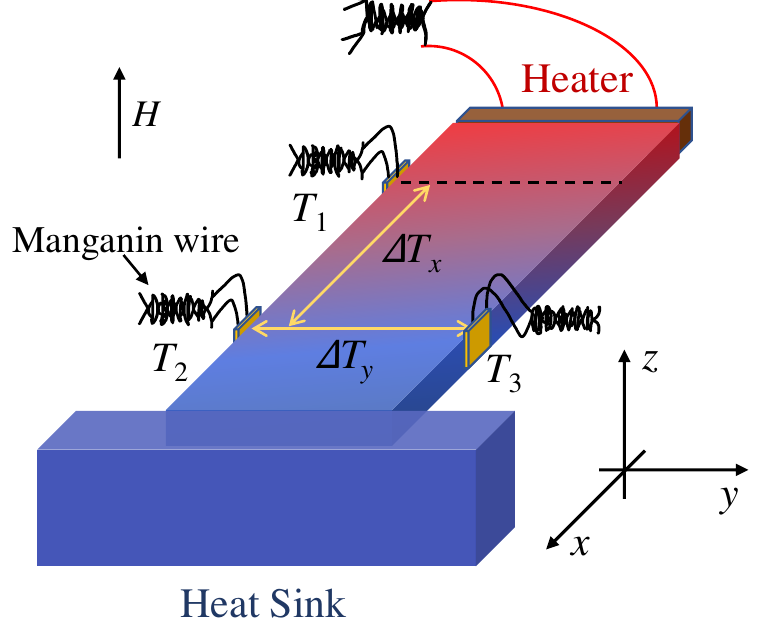} 
		\caption{\textbf{Experimental configuration for measurement of longitudinal/transverse thermal transport. }The magnetic field is applied along the z direction, perpendicular to the sample surface.}
		\label{FIG. S2}
	\end{figure}

	The high‑purity samples used in this work are a commercial cold‑rolled polycrystalline Nb sheet(used in the main text, 5$\times$3.6$\times$0.083 mm$^3$)  and a [110]‑oriented single‑crystalline Nb sheet, cut, polished and annealed from a single‑crystalline bulk(used in
	the Supplementary Materials, 5$\times$3$\times$0.11 mm$^3$). The SEM image of the sample shows elongated grains with a width on the order of 10 $\mu m$, resulting from the cold‑rolling process of the polycrystalline sheet[see Fig.~\ref{FIG. S1}]. This size is comparable to the maximum phonon mean free path obtained in the main text. 
	The electrical transport experiments were performed in a commercial measurement system (Quantum Design PPMS), and the thermal transport experiments were performed in a Leiden dilution refrigerator. The voltage was monitored by DC nanometers (Keithley 2182A)and the electric current was driven by a current source (Keithley 6221). The temperature signals were acquired with three SR830 lock‑in amplifiers. The one-heater
	three-thermometers(Cernox 1030) method was used to simultaneously measure the longitudinal and transverse thermal gradient. The thermal gradient in the sample was produced through a 10 k$\Omega$ chip resistor.  Thermal isolation between the heater/thermometers and the sample holder is achieved by using manganin wires as connection leads[see Fig.~\ref{FIG. S2}]. All contacts on the sample were made using silver paste.
	
	\begin{figure}[ht]
		\centering
		\includegraphics[width=0.45\linewidth]{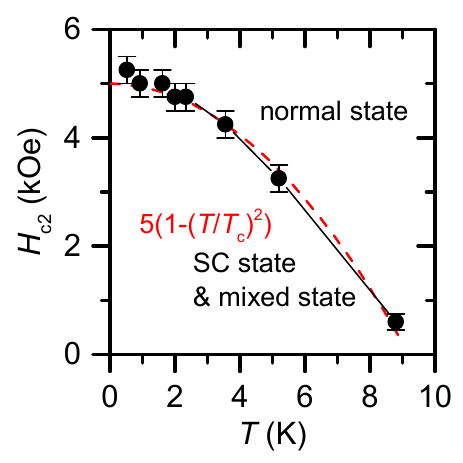} 
		\caption{\textbf{Phase diagram of Niobium. }The red dashed line shows the empirical $H_\mathrm{c2}-T$
			relation.}
		\label{FIG. S3}
	\end{figure}
	
	\section{Field‑cooling Protocol and Field Calibration}
	
	During a field sweep, flux pinning leads to non-uniform flux entry and exit, thereby introducing substantial noise into the thermal Hall signal upon antisymmetrization. Therefore, before acquiring data at each magnetic field point, we applied a strong heating pulse to drive the sample above $T_\mathrm{c}$. This allowed magnetic flux lines to uniformly penetrate the sample in the normal state. The sample was then cooled down to the target temperature for measurement. Given the pronounced sensitivity of the sample to the applied magnetic field, a graphite chip was placed alongside the sample. Using the magnetoresistance effect of graphite, the magnetic field values were calibrated to compensate for deviations arising from the remnant field of the superconducting magnet. Measurements were formed under a fixed background temperature and a constant heat flow. Because the thermal conductivity and specific heat of the sample vary dramatically, the sample temperature inevitably deviates as the magnetic field is changed. The temperature values shown in Fig. 2 are those at zero field. But below 1000 Oe, the sample temperature remains relatively stable, with a maximum fluctuations of about ±0.1 K. The phase diagram, which is constructed from the saturation magnetic field of $\kappa_{xx}$ in Fig. 2 of the main text, is displayed in Fig.~\ref{FIG. S3}. Due to the use of the field‑cooling method, the lower critical field $H_\mathrm{c1}$ is not defined in this plot. The temperature dependence of $H_\mathrm{c2}$ is consistent with the empirical parabolic form.
	
	\begin{figure*}[ht]
		\centering
		\includegraphics[width=1\linewidth]{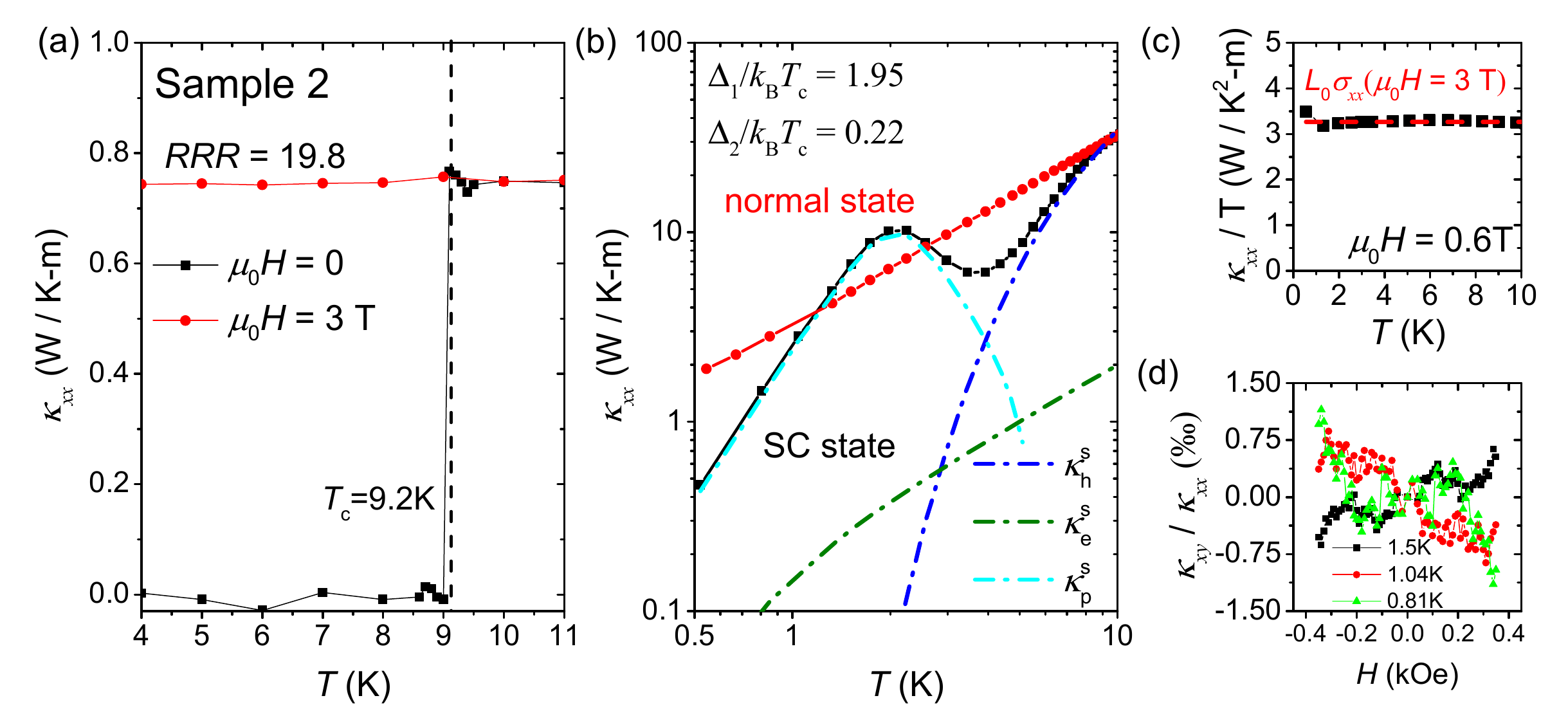} 
		\caption{\textbf{Transport properties in the single crystal sample. }(a) Longitudinal resistivity of single‑crystalline niobium. (b) Longitudinal thermal conductivity in the normal and superconducting states, with the three resolved superconducting components. $\kappa_h^s$ and $\kappa_e^s$ are from BRT fits; $\kappa_p^s$ is obtained by subtracting these from the measured total. Near 0.2$T_\mathrm{c}$ ($\sim$1.8 K), the weakened electron‑phonon scattering allows the phonon thermal conductivity to dominate, even exceeding the normal‑state thermal conductivity of both holes and electrons. (c) $\kappa_{xx}/T$ in the normal state, which satisfies the Wiedemann–Franz law. (d) Thermal Hall angle at various temperatures. At 1.5 K, where phonons are the dominant heat carriers, no sign reversal of the thermal Hall coefficient, as reported in the main text, is observed.}
		\label{FIG. S4}
	\end{figure*}
	
	\section{Thermal Transport in the Single Crystal Sample}
	
	In order to ascertain whether the sign reversal of the thermal Hall coefficient reported in the main text is of phononic origin, we carried out measurements on a single‑crystalline niobium sample. The longitudinal electrical and thermal transport data are presented in Figs.~\ref{FIG. S4}(a, c). The $T_\mathrm{c}$ agrees with that of the polycrystal in the main text, and the Wiedemann–Franz law is well satisfied. Compared with its polycrystalline counterpart, the single crystal is free of grain‑boundary scattering, allowing a longer mean free path of phonons and consequently a higher phonon thermal conductivity. However, the altered scattering mechanism increases the difficulty in quantitatively evaluating $\kappa_p^s$ (whereas in the polycrystal, the mean free path of phonons easily saturates at grain size and becomes constant). A fit to this single‑crystalline sample is performed using the gap magnitudes and the hole‑to‑electron ratio obtained from the main text; the results are presented in Fig.~\ref{FIG. S4}(b). At some temperatures (e.g., 1.5 K, corresponding to about 0.16$T_\mathrm{c}$), $\kappa_p^s$ becomes absolutely dominant, even exceeding the hole/electron thermal conductivity in the normal metallic state. However, its thermal Hall angle does not exhibit a response opposite to that of holes. This further validates the view that the electron band is associated with a second energy gap.
	
	\begin{figure}[ht]
		\centering
		\includegraphics[width=1\linewidth]{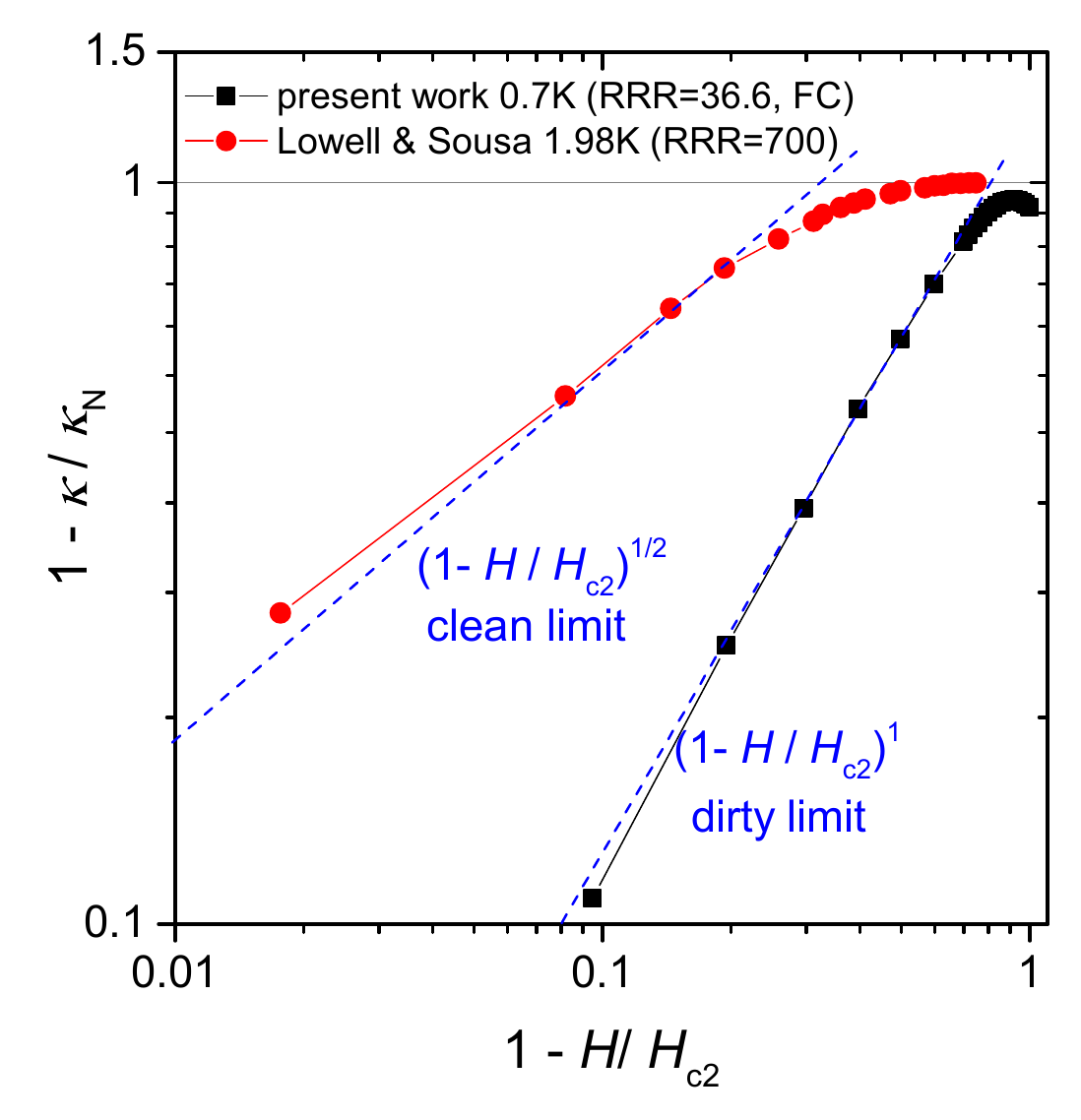} 
		\caption{\textbf{The field dependence of thermal conductivity in the vicinity of the upper critical field. }The dashed curve represents the power‑law relation predicted by theory.}
		\label{FIG. S5}
	\end{figure}
	
	\section{Field Dependence of Thermal Conductivity} 
	
	In the zero‑temperature limit, the linear term of thermal conductivity $\kappa/T$ of a fully gapped superconductor approaches zero ($\kappa_0/T\to 0$), since the presence of the energy gap prevents any quasiparticle excitations. Magnetic fields can excite quasiparticles in type‑II superconductors even at zero temperature, serving as a probe of the finite‑energy excitation spectrum. For an s‑wave superconductor when $T \to 0$, quasiparticles are confined to the vortex core states, and any thermal transport perpendicular to the field arises from inter‑vortex tunneling. As shown in Fig.~\ref{FIG. S5}, in the vicinity of the upper critical field ($H \to H_\mathrm{c2}$), our data follow $(1-\kappa/\kappa_n)\propto (1-H/H_\mathrm{c2})$, which is consistent with the prediction of Caroli and Cyrot for the dirty limit \cite{caroli1965}. The data of Lowell and Sousa, on the other hand, satisfy $(1-\kappa/\kappa_n)\propto (1-H/H_\mathrm{c2})^{1/2}$, which is consistent with Maki's prediction for the clean limit \cite{Maki1967}. 
	
	\begin{figure}[ht]
		\centering
		\includegraphics[width=0.85\linewidth]{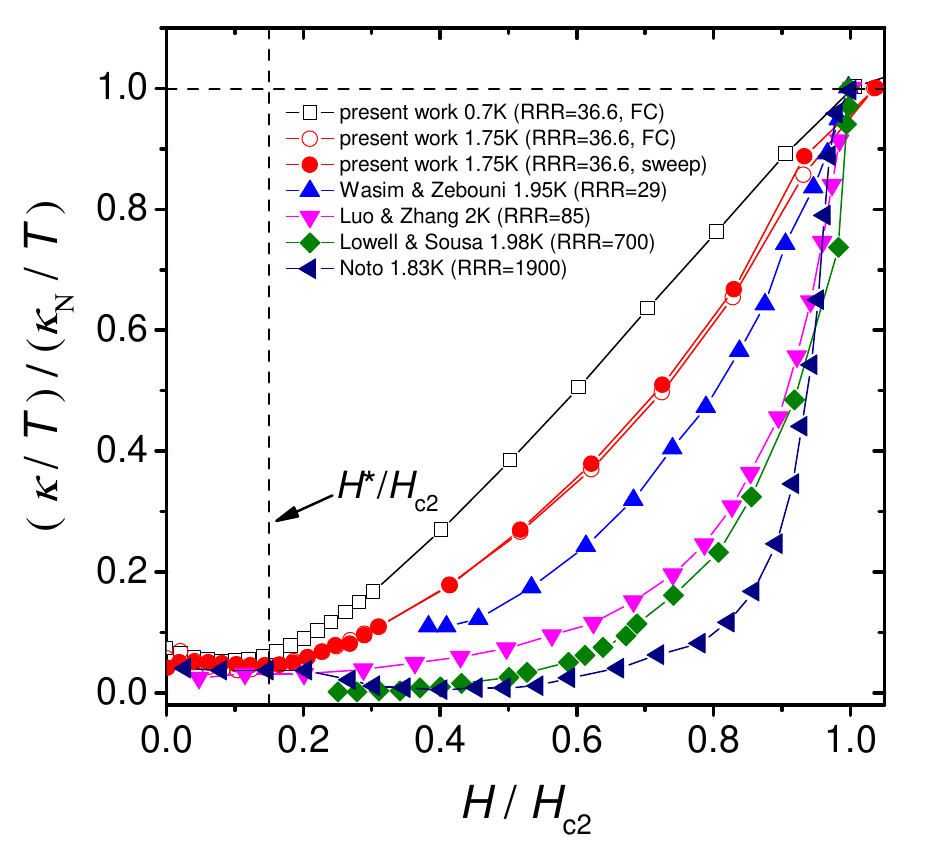} 
		\caption{\textbf{The normalized field dependence of thermal conductivity. }The vertical dashed line marks the expected position of $H^*/H_\mathrm{c2}$. }
		\label{FIG. S6}
	\end{figure}
	
	Fig.~\ref{FIG. S6} shows the normalized field dependence of thermal conductivity in our work and in previous studies \cite{Wasim1969,Luo2025,Lowell1970,noto1969}. The difference in curve positions originates from variations in temperature and sample quality, while the field‑cooling measurement method does not introduce any appreciable change (see red open circles and solid circles). In two-gap s-wave superconductors
	whose gap is very different on two parts of the Fermi surface, since $H_\mathrm{c2} \propto \Delta^2 / v_F^2$, each Fermi surface has its own $H_\mathrm{c2}$. This appears as a shoulder in the $\kappa_0/T$ versus $H$ curve at $H^*=[\frac{\Delta_2v_{F1}}{\Delta_1v_{F2}}]^2 H_\mathrm{c2}$, seen in both $\mathrm{MgB}_2$ \cite{sologubenko2002} and $\mathrm{NbSe}_2$ \cite{Boaknin2003}. Since this feature was not observed in earlier studies on niobium, it had long been regarded a prototypical single‑gap s‑wave superconductor. Assuming that the effective masses of holes and electrons are not significantly different, we have $v_F \propto n^{1/3}$. Combining this with the fitted magnitudes of $\Delta_1$ and $\Delta_2$, the predicted $H^*$ is only about 15\% of $H_\mathrm{c2}$, which is even lower than $H_\mathrm{c1}$, as shown in  Fig.~\ref{FIG. S6}. For the conventional field‑sweep method, the magnetic field cannot penetrate the sample at all because $H^* < H_\mathrm{c1}$. Although the field‑cooling protocol employed in our work permits magnetic flux to enter the sample, below this field ($\sim$ 750 Oe) the thermal conductivity is severely modified by phonon scattering from vortices (see Fig. 2 in the main text). Thus, the evolution of quasiparticle properties is difficult to detect. This inaccessible measurement regime is exactly what prevented previous magnetothermal conductivity experiments from detecting two‑gap signatures. 
	
	\section{Re-fitting of Literature $\kappa(T)$ Data} 
	
\begin{figure}[ht]
		\centering
		\includegraphics[width=0.85\linewidth]{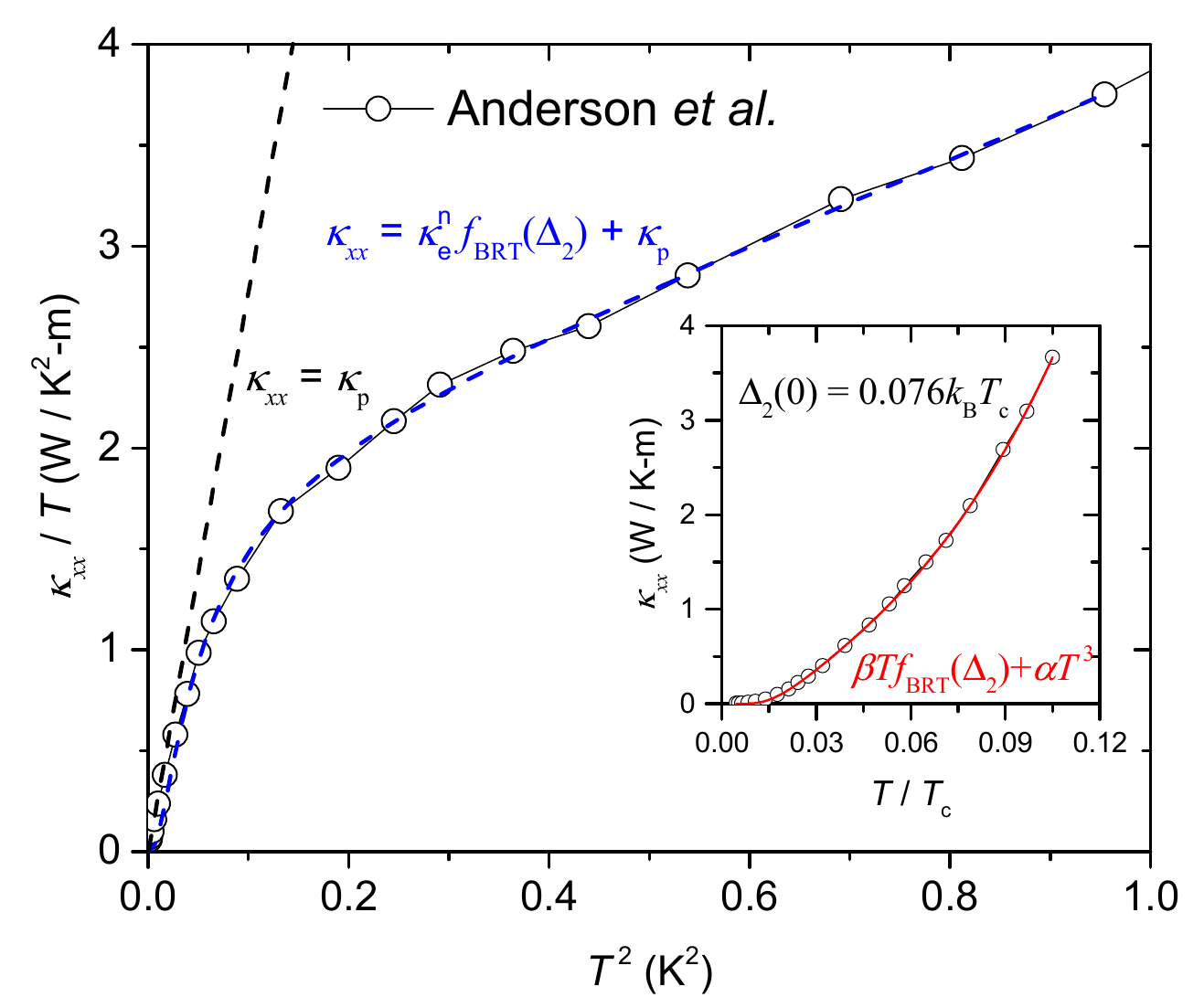} 
		\caption{\textbf{Two‑gap BRT model fit at low temperatures. }The black solid line is the  $T^3$ prediction for ballistic phonon transport given by Anderson \textit{et al.} The blue dashed line and the red solid line are the fitting curves obtained by our two‑gap model.}
		\label{FIG. S7}
	\end{figure}

	Following Carlson's 1970 thermal conductivity evidence for the two-‐gap model \cite{carlson1970anomalous}, Anderson soon asserted that his own data did not show this \cite{Anderson1971}. Here we reanalyze and fit Anderson's data. Fig.~\ref{FIG. S7} shows Anderson's predicted line for ballistic phonon transport at low temperatures (black dashed line). He stated that phonons become ballistic only below about 0.1K, and no fit was given for the region above that temperature. We argue that Anderson significantly overestimated the phonon mean free path in his sample. In fact, phonons enter the ballistic transport regime already below 1K, and The reason why $\kappa_{xx}/T$ is not proportional to $T^2$ is the additional quasiparticle excitations due to the second energy gap. The blue and red dashed curves in Fig.~\ref{FIG. S7} show that the model provides an excellent fit, giving a second gap of 
	0.076 $k_\mathrm{B}T_\mathrm{c}$
	, very close to Carlson's value of 0.07$k_\mathrm{B}T_\mathrm{c}$.
	
	\clearpage

\begin{thebibliography}{67}%
	\makeatletter
	\providecommand \@ifxundefined [1]{%
		\@ifx{#1\undefined}
	}%
	\providecommand \@ifnum [1]{%
		\ifnum #1\expandafter \@firstoftwo
		\else \expandafter \@secondoftwo
		\fi
	}%
	\providecommand \@ifx [1]{%
		\ifx #1\expandafter \@firstoftwo
		\else \expandafter \@secondoftwo
		\fi
	}%
	\providecommand \natexlab [1]{#1}%
	\providecommand \enquote  [1]{``#1''}%
	\providecommand \bibnamefont  [1]{#1}%
	\providecommand \bibfnamefont [1]{#1}%
	\providecommand \citenamefont [1]{#1}%
	\providecommand \href@noop [0]{\@secondoftwo}%
	\providecommand \href [0]{\begingroup \@sanitize@url \@href}%
	\providecommand \@href[1]{\@@startlink{#1}\@@href}%
	\providecommand \@@href[1]{\endgroup#1\@@endlink}%
	\providecommand \@sanitize@url [0]{\catcode `\\12\catcode `\$12\catcode
		`\&12\catcode `\#12\catcode `\^12\catcode `\_12\catcode `\%12\relax}%
	\providecommand \@@startlink[1]{}%
	\providecommand \@@endlink[0]{}%
	\providecommand \url  [0]{\begingroup\@sanitize@url \@url }%
	\providecommand \@url [1]{\endgroup\@href {#1}{\urlprefix }}%
	\providecommand \urlprefix  [0]{URL }%
	\providecommand \Eprint [0]{\href }%
	\providecommand \doibase [0]{https://doi.org/}%
	\providecommand \selectlanguage [0]{\@gobble}%
	\providecommand \bibinfo  [0]{\@secondoftwo}%
	\providecommand \bibfield  [0]{\@secondoftwo}%
	\providecommand \translation [1]{[#1]}%
	\providecommand \BibitemOpen [0]{}%
	\providecommand \bibitemStop [0]{}%
	\providecommand \bibitemNoStop [0]{.\EOS\space}%
	\providecommand \EOS [0]{\spacefactor3000\relax}%
	\providecommand \BibitemShut  [1]{\csname bibitem#1\endcsname}%
	\let\auto@bib@innerbib\@empty
	\bibitem [{\citenamefont {Bardeen}\ \emph {et~al.}(1957)\citenamefont
		{Bardeen}, \citenamefont {Cooper},\ and\ \citenamefont
		{Schrieffer}}]{Bardeen1957}%
	\BibitemOpen
	\bibfield  {author} {\bibinfo {author} {\bibfnamefont {J.}~\bibnamefont
			{Bardeen}}, \bibinfo {author} {\bibfnamefont {L.~N.}\ \bibnamefont
			{Cooper}},\ and\ \bibinfo {author} {\bibfnamefont {J.~R.}\ \bibnamefont
			{Schrieffer}},\ }\bibfield  {title} {\bibinfo {title} {Theory of
			superconductivity},\ }\href {https://doi.org/10.1103/PhysRev.108.1175}
	{\bibfield  {journal} {\bibinfo  {journal} {Phys. Rev.}\ }\textbf {\bibinfo
			{volume} {108}},\ \bibinfo {pages} {1175} (\bibinfo {year}
		{1957})}\BibitemShut {NoStop}%
	\bibitem [{\citenamefont {Suhl}\ \emph {et~al.}(1959)\citenamefont {Suhl},
		\citenamefont {Matthias},\ and\ \citenamefont {Walker}}]{Suhl1959}%
	\BibitemOpen
	\bibfield  {author} {\bibinfo {author} {\bibfnamefont {H.}~\bibnamefont
			{Suhl}}, \bibinfo {author} {\bibfnamefont {B.~T.}\ \bibnamefont {Matthias}},\
		and\ \bibinfo {author} {\bibfnamefont {L.~R.}\ \bibnamefont {Walker}},\
	}\bibfield  {title} {\bibinfo {title} {{Bardeen-Cooper-Schrieffer} theory of
			superconductivity in the case of overlapping bands},\ }\href
	{https://doi.org/10.1103/PhysRevLett.3.552} {\bibfield  {journal} {\bibinfo
			{journal} {Phys. Rev. Lett.}\ }\textbf {\bibinfo {volume} {3}},\ \bibinfo
		{pages} {552} (\bibinfo {year} {1959})}\BibitemShut {NoStop}%
	\bibitem [{\citenamefont {Matthias}(1955)}]{Matthias1955}%
	\BibitemOpen
	\bibfield  {author} {\bibinfo {author} {\bibfnamefont {B.~T.}\ \bibnamefont
			{Matthias}},\ }\bibfield  {title} {\bibinfo {title} {Empirical relation
			between superconductivity and the number of valence electrons per atom},\
	}\href {https://doi.org/10.1103/PhysRev.97.74} {\bibfield  {journal}
		{\bibinfo  {journal} {Physical Review}\ }\textbf {\bibinfo {volume} {97}},\
		\bibinfo {pages} {74} (\bibinfo {year} {1955})}\BibitemShut {NoStop}%
	\bibitem [{\citenamefont {MacVicar}\ and\ \citenamefont
		{Rose}(1968)}]{macvicar1968}%
	\BibitemOpen
	\bibfield  {author} {\bibinfo {author} {\bibfnamefont {M.}~\bibnamefont
			{MacVicar}}\ and\ \bibinfo {author} {\bibfnamefont {R.}~\bibnamefont
			{Rose}},\ }\bibfield  {title} {\bibinfo {title} {Anisotropic energy-gap
			measurements on superconducting niobium single crystals by tunneling},\
	}\href {https://doi.org/10.1063/1.1656421} {\bibfield  {journal} {\bibinfo
			{journal} {Journal of Applied Physics}\ }\textbf {\bibinfo {volume} {39}},\
		\bibinfo {pages} {1721} (\bibinfo {year} {1968})}\BibitemShut {NoStop}%
	\bibitem [{\citenamefont {Hess}\ \emph {et~al.}(1991)\citenamefont {Hess},
		\citenamefont {Robinson},\ and\ \citenamefont {Waszczak}}]{HESS1991422}%
	\BibitemOpen
	\bibfield  {author} {\bibinfo {author} {\bibfnamefont {H.}~\bibnamefont
			{Hess}}, \bibinfo {author} {\bibfnamefont {R.}~\bibnamefont {Robinson}},\
		and\ \bibinfo {author} {\bibfnamefont {J.}~\bibnamefont {Waszczak}},\
	}\bibfield  {title} {\bibinfo {title} {{STM} spectroscopy of vortex cores and
			the flux lattice},\ }\href
	{https://doi.org/https://doi.org/10.1016/0921-4526(91)90262-D} {\bibfield
		{journal} {\bibinfo  {journal} {Physica B: Condensed Matter}\ }\textbf
		{\bibinfo {volume} {169}},\ \bibinfo {pages} {422} (\bibinfo {year}
		{1991})}\BibitemShut {NoStop}%
	\bibitem [{\citenamefont {Hahn}\ \emph {et~al.}(1998)\citenamefont {Hahn},
		\citenamefont {Hofmann}, \citenamefont {Krause},\ and\ \citenamefont
		{Seidel}}]{HAHN1998}%
	\BibitemOpen
	\bibfield  {author} {\bibinfo {author} {\bibfnamefont {A.}~\bibnamefont
			{Hahn}}, \bibinfo {author} {\bibfnamefont {S.}~\bibnamefont {Hofmann}},
		\bibinfo {author} {\bibfnamefont {A.}~\bibnamefont {Krause}},\ and\ \bibinfo
		{author} {\bibfnamefont {P.}~\bibnamefont {Seidel}},\ }\bibfield  {title}
	{\bibinfo {title} {Tunneling results on gap anisotropy in niobium},\ }\href
	{https://doi.org/https://doi.org/10.1016/S0921-4534(97)01826-1} {\bibfield
		{journal} {\bibinfo  {journal} {Physica C: Superconductivity}\ }\textbf
		{\bibinfo {volume} {296}},\ \bibinfo {pages} {103} (\bibinfo {year}
		{1998})}\BibitemShut {NoStop}%
	\bibitem [{\citenamefont {Boaknin}\ \emph {et~al.}(2003)\citenamefont
		{Boaknin}, \citenamefont {Tanatar}, \citenamefont {Paglione}, \citenamefont
		{Hawthorn}, \citenamefont {Ronning}, \citenamefont {Hill}, \citenamefont
		{Sutherland}, \citenamefont {Taillefer}, \citenamefont {Sonier},
		\citenamefont {Hayden},\ and\ \citenamefont {Brill}}]{Boaknin2003}%
	\BibitemOpen
	\bibfield  {author} {\bibinfo {author} {\bibfnamefont {E.}~\bibnamefont
			{Boaknin}}, \bibinfo {author} {\bibfnamefont {M.~A.}\ \bibnamefont
			{Tanatar}}, \bibinfo {author} {\bibfnamefont {J.}~\bibnamefont {Paglione}},
		\bibinfo {author} {\bibfnamefont {D.}~\bibnamefont {Hawthorn}}, \bibinfo
		{author} {\bibfnamefont {F.}~\bibnamefont {Ronning}}, \bibinfo {author}
		{\bibfnamefont {R.~W.}\ \bibnamefont {Hill}}, \bibinfo {author}
		{\bibfnamefont {M.}~\bibnamefont {Sutherland}}, \bibinfo {author}
		{\bibfnamefont {L.}~\bibnamefont {Taillefer}}, \bibinfo {author}
		{\bibfnamefont {J.}~\bibnamefont {Sonier}}, \bibinfo {author} {\bibfnamefont
			{S.~M.}\ \bibnamefont {Hayden}},\ and\ \bibinfo {author} {\bibfnamefont
			{J.~W.}\ \bibnamefont {Brill}},\ }\bibfield  {title} {\bibinfo {title} {Heat
			conduction in the vortex state of $\mathrm{NbSe_2}$: Evidence for multiband
			superconductivity},\ }\href {https://doi.org/10.1103/PhysRevLett.90.117003}
	{\bibfield  {journal} {\bibinfo  {journal} {Phys. Rev. Lett.}\ }\textbf
		{\bibinfo {volume} {90}},\ \bibinfo {pages} {117003} (\bibinfo {year}
		{2003})}\BibitemShut {NoStop}%
	\bibitem [{\citenamefont {Rodrigo}\ and\ \citenamefont
		{Vieira}(2004)}]{RODRIGO2004306}%
	\BibitemOpen
	\bibfield  {author} {\bibinfo {author} {\bibfnamefont {J.}~\bibnamefont
			{Rodrigo}}\ and\ \bibinfo {author} {\bibfnamefont {S.}~\bibnamefont
			{Vieira}},\ }\bibfield  {title} {\bibinfo {title} {{STM} study of multiband
			superconductivity in $\mathrm{NbSe_2}$ using a superconducting tip},\ }\href
	{https://doi.org/https://doi.org/10.1016/j.physc.2003.10.030} {\bibfield
		{journal} {\bibinfo  {journal} {Physica C: Superconductivity}\ }\textbf
		{\bibinfo {volume} {404}},\ \bibinfo {pages} {306} (\bibinfo {year}
		{2004})}\BibitemShut {NoStop}%
	\bibitem [{\citenamefont {Fletcher}\ \emph {et~al.}(2007)\citenamefont
		{Fletcher}, \citenamefont {Carrington}, \citenamefont {Diener}, \citenamefont
		{Rodi\`ere}, \citenamefont {Brison}, \citenamefont {Prozorov}, \citenamefont
		{Olheiser},\ and\ \citenamefont {Giannetta}}]{Fletcher2007}%
	\BibitemOpen
	\bibfield  {author} {\bibinfo {author} {\bibfnamefont {J.~D.}\ \bibnamefont
			{Fletcher}}, \bibinfo {author} {\bibfnamefont {A.}~\bibnamefont
			{Carrington}}, \bibinfo {author} {\bibfnamefont {P.}~\bibnamefont {Diener}},
		\bibinfo {author} {\bibfnamefont {P.}~\bibnamefont {Rodi\`ere}}, \bibinfo
		{author} {\bibfnamefont {J.~P.}\ \bibnamefont {Brison}}, \bibinfo {author}
		{\bibfnamefont {R.}~\bibnamefont {Prozorov}}, \bibinfo {author}
		{\bibfnamefont {T.}~\bibnamefont {Olheiser}},\ and\ \bibinfo {author}
		{\bibfnamefont {R.~W.}\ \bibnamefont {Giannetta}},\ }\bibfield  {title}
	{\bibinfo {title} {Penetration depth study of superconducting gap structure
			of $\mathrm{2H\text{-}NbSe_2}$},\ }\href
	{https://doi.org/10.1103/PhysRevLett.98.057003} {\bibfield  {journal}
		{\bibinfo  {journal} {Phys. Rev. Lett.}\ }\textbf {\bibinfo {volume} {98}},\
		\bibinfo {pages} {057003} (\bibinfo {year} {2007})}\BibitemShut {NoStop}%
	\bibitem [{\citenamefont {Zehetmayer}\ and\ \citenamefont
		{Weber}(2010)}]{Zehetmayer2010}%
	\BibitemOpen
	\bibfield  {author} {\bibinfo {author} {\bibfnamefont {M.}~\bibnamefont
			{Zehetmayer}}\ and\ \bibinfo {author} {\bibfnamefont {H.~W.}\ \bibnamefont
			{Weber}},\ }\bibfield  {title} {\bibinfo {title} {Experimental evidence for a
			two-band superconducting state of $\mathrm{NbSe_2}$ single crystals},\ }\href
	{https://doi.org/10.1103/PhysRevB.82.014524} {\bibfield  {journal} {\bibinfo
			{journal} {Phys. Rev. B}\ }\textbf {\bibinfo {volume} {82}},\ \bibinfo
		{pages} {014524} (\bibinfo {year} {2010})}\BibitemShut {NoStop}%
	\bibitem [{\citenamefont {Sanna}\ \emph {et~al.}(2022)\citenamefont {Sanna},
		\citenamefont {Pellegrini}, \citenamefont {Liebhaber}, \citenamefont
		{Rossnagel}, \citenamefont {Franke},\ and\ \citenamefont
		{Gross}}]{sanna2022}%
	\BibitemOpen
	\bibfield  {author} {\bibinfo {author} {\bibfnamefont {A.}~\bibnamefont
			{Sanna}}, \bibinfo {author} {\bibfnamefont {C.}~\bibnamefont {Pellegrini}},
		\bibinfo {author} {\bibfnamefont {E.}~\bibnamefont {Liebhaber}}, \bibinfo
		{author} {\bibfnamefont {K.}~\bibnamefont {Rossnagel}}, \bibinfo {author}
		{\bibfnamefont {K.~J.}\ \bibnamefont {Franke}},\ and\ \bibinfo {author}
		{\bibfnamefont {E.}~\bibnamefont {Gross}},\ }\bibfield  {title} {\bibinfo
		{title} {Real-space anisotropy of the superconducting gap in the
			charge-density wave material $\mathrm{2H\text{-}NbSe_2}$},\ }\href
	{https://doi.org/10.1038/s41535-021-00412-8} {\bibfield  {journal} {\bibinfo
			{journal} {npj Quantum Materials}\ }\textbf {\bibinfo {volume} {7}},\
		\bibinfo {pages} {6} (\bibinfo {year} {2022})}\BibitemShut {NoStop}%
	\bibitem [{\citenamefont {Alshemi}\ \emph {et~al.}(2025)\citenamefont
		{Alshemi}, \citenamefont {Forgan}, \citenamefont {Hiess}, \citenamefont
		{Cubitt}, \citenamefont {White}, \citenamefont {Schmalzl},\ and\
		\citenamefont {Blackburn}}]{Alshemi2025}%
	\BibitemOpen
	\bibfield  {author} {\bibinfo {author} {\bibfnamefont {A.}~\bibnamefont
			{Alshemi}}, \bibinfo {author} {\bibfnamefont {E.~M.}\ \bibnamefont {Forgan}},
		\bibinfo {author} {\bibfnamefont {A.}~\bibnamefont {Hiess}}, \bibinfo
		{author} {\bibfnamefont {R.}~\bibnamefont {Cubitt}}, \bibinfo {author}
		{\bibfnamefont {J.~S.}\ \bibnamefont {White}}, \bibinfo {author}
		{\bibfnamefont {K.}~\bibnamefont {Schmalzl}},\ and\ \bibinfo {author}
		{\bibfnamefont {E.}~\bibnamefont {Blackburn}},\ }\bibfield  {title} {\bibinfo
		{title} {Two characteristic contributions to the superconducting state of
			$\mathrm{2H\text{-}NbSe_2}$},\ }\href
	{https://doi.org/10.1103/PhysRevLett.134.116001} {\bibfield  {journal}
		{\bibinfo  {journal} {Phys. Rev. Lett.}\ }\textbf {\bibinfo {volume} {134}},\
		\bibinfo {pages} {116001} (\bibinfo {year} {2025})}\BibitemShut {NoStop}%
	\bibitem [{\citenamefont {Ruby}\ \emph {et~al.}(2015)\citenamefont {Ruby},
		\citenamefont {Heinrich}, \citenamefont {Pascual},\ and\ \citenamefont
		{Franke}}]{Ruby2015}%
	\BibitemOpen
	\bibfield  {author} {\bibinfo {author} {\bibfnamefont {M.}~\bibnamefont
			{Ruby}}, \bibinfo {author} {\bibfnamefont {B.~W.}\ \bibnamefont {Heinrich}},
		\bibinfo {author} {\bibfnamefont {J.~I.}\ \bibnamefont {Pascual}},\ and\
		\bibinfo {author} {\bibfnamefont {K.~J.}\ \bibnamefont {Franke}},\ }\bibfield
	{title} {\bibinfo {title} {Experimental demonstration of a two-band
			superconducting state for lead using scanning tunneling spectroscopy},\
	}\href {https://doi.org/10.1103/PhysRevLett.114.157001} {\bibfield  {journal}
		{\bibinfo  {journal} {Phys. Rev. Lett.}\ }\textbf {\bibinfo {volume} {114}},\
		\bibinfo {pages} {157001} (\bibinfo {year} {2015})}\BibitemShut {NoStop}%
	\bibitem [{\citenamefont {Khasanov}\ \emph {et~al.}(2021)\citenamefont
		{Khasanov}, \citenamefont {Das}, \citenamefont {Gawryluk}, \citenamefont
		{Gupta},\ and\ \citenamefont {Mielke~III}}]{khasanov2021}%
	\BibitemOpen
	\bibfield  {author} {\bibinfo {author} {\bibfnamefont {R.}~\bibnamefont
			{Khasanov}}, \bibinfo {author} {\bibfnamefont {D.}~\bibnamefont {Das}},
		\bibinfo {author} {\bibfnamefont {D.~J.}\ \bibnamefont {Gawryluk}}, \bibinfo
		{author} {\bibfnamefont {R.}~\bibnamefont {Gupta}},\ and\ \bibinfo {author}
		{\bibfnamefont {C.}~\bibnamefont {Mielke~III}},\ }\bibfield  {title}
	{\bibinfo {title} {Isotropic single-gap superconductivity of elemental
			{Pb}},\ }\href {https://doi.org/10.1103/PhysRevB.104.L100508} {\bibfield
		{journal} {\bibinfo  {journal} {Phys. Rev. B}\ }\textbf {\bibinfo {volume}
			{104}},\ \bibinfo {pages} {L100508} (\bibinfo {year} {2021})}\BibitemShut
	{NoStop}%
	\bibitem [{\citenamefont {Xu}\ \emph {et~al.}(2016)\citenamefont {Xu},
		\citenamefont {Niu}, \citenamefont {Xu}, \citenamefont {Jiang}, \citenamefont
		{Yao}, \citenamefont {Chen}, \citenamefont {Song}, \citenamefont
		{Abdel-Hafiez}, \citenamefont {Chareev}, \citenamefont {Vasiliev},
		\citenamefont {Wang}, \citenamefont {Wo}, \citenamefont {Zhao}, \citenamefont
		{Peng},\ and\ \citenamefont {Feng}}]{Xu2016}%
	\BibitemOpen
	\bibfield  {author} {\bibinfo {author} {\bibfnamefont {H.~C.}\ \bibnamefont
			{Xu}}, \bibinfo {author} {\bibfnamefont {X.~H.}\ \bibnamefont {Niu}},
		\bibinfo {author} {\bibfnamefont {D.~F.}\ \bibnamefont {Xu}}, \bibinfo
		{author} {\bibfnamefont {J.}~\bibnamefont {Jiang}}, \bibinfo {author}
		{\bibfnamefont {Q.}~\bibnamefont {Yao}}, \bibinfo {author} {\bibfnamefont
			{Q.~Y.}\ \bibnamefont {Chen}}, \bibinfo {author} {\bibfnamefont
			{Q.}~\bibnamefont {Song}}, \bibinfo {author} {\bibfnamefont {M.}~\bibnamefont
			{Abdel-Hafiez}}, \bibinfo {author} {\bibfnamefont {D.~A.}\ \bibnamefont
			{Chareev}}, \bibinfo {author} {\bibfnamefont {A.~N.}\ \bibnamefont
			{Vasiliev}}, \bibinfo {author} {\bibfnamefont {Q.~S.}\ \bibnamefont {Wang}},
		\bibinfo {author} {\bibfnamefont {H.~L.}\ \bibnamefont {Wo}}, \bibinfo
		{author} {\bibfnamefont {J.}~\bibnamefont {Zhao}}, \bibinfo {author}
		{\bibfnamefont {R.}~\bibnamefont {Peng}},\ and\ \bibinfo {author}
		{\bibfnamefont {D.~L.}\ \bibnamefont {Feng}},\ }\bibfield  {title} {\bibinfo
		{title} {Highly anisotropic and twofold symmetric superconducting gap in
			nematically ordered $\mathrm{FeSe_{0.93} S_{0.07}}$},\ }\href
	{https://doi.org/10.1103/PhysRevLett.117.157003} {\bibfield  {journal}
		{\bibinfo  {journal} {Phys. Rev. Lett.}\ }\textbf {\bibinfo {volume} {117}},\
		\bibinfo {pages} {157003} (\bibinfo {year} {2016})}\BibitemShut {NoStop}%
	\bibitem [{\citenamefont {Hashimoto}\ \emph {et~al.}(2018)\citenamefont
		{Hashimoto}, \citenamefont {Ota}, \citenamefont {Yamamoto}, \citenamefont
		{Suzuki}, \citenamefont {Shimojima}, \citenamefont {Watanabe}, \citenamefont
		{Chen}, \citenamefont {Kasahara}, \citenamefont {Matsuda}, \citenamefont
		{Shibauchi} \emph {et~al.}}]{hashimoto2018}%
	\BibitemOpen
	\bibfield  {author} {\bibinfo {author} {\bibfnamefont {T.}~\bibnamefont
			{Hashimoto}}, \bibinfo {author} {\bibfnamefont {Y.}~\bibnamefont {Ota}},
		\bibinfo {author} {\bibfnamefont {H.~Q.}\ \bibnamefont {Yamamoto}}, \bibinfo
		{author} {\bibfnamefont {Y.}~\bibnamefont {Suzuki}}, \bibinfo {author}
		{\bibfnamefont {T.}~\bibnamefont {Shimojima}}, \bibinfo {author}
		{\bibfnamefont {S.}~\bibnamefont {Watanabe}}, \bibinfo {author}
		{\bibfnamefont {C.}~\bibnamefont {Chen}}, \bibinfo {author} {\bibfnamefont
			{S.}~\bibnamefont {Kasahara}}, \bibinfo {author} {\bibfnamefont
			{Y.}~\bibnamefont {Matsuda}}, \bibinfo {author} {\bibfnamefont
			{T.}~\bibnamefont {Shibauchi}}, \emph {et~al.},\ }\bibfield  {title}
	{\bibinfo {title} {Superconducting gap anisotropy sensitive to nematic
			domains in {FeSe}},\ }\href {https://doi.org/10.1038/s41467-017-02739-y}
	{\bibfield  {journal} {\bibinfo  {journal} {Nature communications}\ }\textbf
		{\bibinfo {volume} {9}},\ \bibinfo {pages} {282} (\bibinfo {year}
		{2018})}\BibitemShut {NoStop}%
	\bibitem [{\citenamefont {Sun}\ \emph {et~al.}(2018)\citenamefont {Sun},
		\citenamefont {Kittaka}, \citenamefont {Nakamura}, \citenamefont
		{Sakakibara}, \citenamefont {Zhang}, \citenamefont {Shin}, \citenamefont
		{Irie}, \citenamefont {Nomoto}, \citenamefont {Machida}, \citenamefont {Chen}
		\emph {et~al.}}]{sun2018}%
	\BibitemOpen
	\bibfield  {author} {\bibinfo {author} {\bibfnamefont {Y.}~\bibnamefont
			{Sun}}, \bibinfo {author} {\bibfnamefont {S.}~\bibnamefont {Kittaka}},
		\bibinfo {author} {\bibfnamefont {S.}~\bibnamefont {Nakamura}}, \bibinfo
		{author} {\bibfnamefont {T.}~\bibnamefont {Sakakibara}}, \bibinfo {author}
		{\bibfnamefont {P.}~\bibnamefont {Zhang}}, \bibinfo {author} {\bibfnamefont
			{S.}~\bibnamefont {Shin}}, \bibinfo {author} {\bibfnamefont {K.}~\bibnamefont
			{Irie}}, \bibinfo {author} {\bibfnamefont {T.}~\bibnamefont {Nomoto}},
		\bibinfo {author} {\bibfnamefont {K.}~\bibnamefont {Machida}}, \bibinfo
		{author} {\bibfnamefont {J.}~\bibnamefont {Chen}}, \emph {et~al.},\
	}\bibfield  {title} {\bibinfo {title} {Disorder-sensitive nodelike small gap
			in {FeSe}},\ }\href {https://doi.org/10.1103/PhysRevB.98.064505} {\bibfield
		{journal} {\bibinfo  {journal} {Phys. Rev. B}\ }\textbf {\bibinfo {volume}
			{98}},\ \bibinfo {pages} {064505} (\bibinfo {year} {2018})}\BibitemShut
	{NoStop}%
	\bibitem [{\citenamefont {Zhao}\ \emph {et~al.}(2024)\citenamefont {Zhao},
		\citenamefont {Cui}, \citenamefont {Liu}, \citenamefont {Gong}, \citenamefont
		{Zhang}, \citenamefont {Jia}, \citenamefont {Zang}, \citenamefont {Hu},
		\citenamefont {Zhang}, \citenamefont {Wang} \emph {et~al.}}]{zhao2024}%
	\BibitemOpen
	\bibfield  {author} {\bibinfo {author} {\bibfnamefont {D.}~\bibnamefont
			{Zhao}}, \bibinfo {author} {\bibfnamefont {W.}~\bibnamefont {Cui}}, \bibinfo
		{author} {\bibfnamefont {Y.}~\bibnamefont {Liu}}, \bibinfo {author}
		{\bibfnamefont {G.}~\bibnamefont {Gong}}, \bibinfo {author} {\bibfnamefont
			{L.}~\bibnamefont {Zhang}}, \bibinfo {author} {\bibfnamefont
			{G.}~\bibnamefont {Jia}}, \bibinfo {author} {\bibfnamefont {Y.}~\bibnamefont
			{Zang}}, \bibinfo {author} {\bibfnamefont {X.}~\bibnamefont {Hu}}, \bibinfo
		{author} {\bibfnamefont {D.}~\bibnamefont {Zhang}}, \bibinfo {author}
		{\bibfnamefont {Y.}~\bibnamefont {Wang}}, \emph {et~al.},\ }\bibfield
	{title} {\bibinfo {title} {Electronic inhomogeneity and phase fluctuation in
			one-unit-cell {FeSe} films},\ }\href
	{https://doi.org/10.1038/s41467-024-47350-0} {\bibfield  {journal} {\bibinfo
			{journal} {Nature Communications}\ }\textbf {\bibinfo {volume} {15}},\
		\bibinfo {pages} {3369} (\bibinfo {year} {2024})}\BibitemShut {NoStop}%
	\bibitem [{\citenamefont {Nag}\ \emph {et~al.}(2025)\citenamefont {Nag},
		\citenamefont {Scott}, \citenamefont {de~Carvalho}, \citenamefont {Byland},
		\citenamefont {Yang}, \citenamefont {Walker}, \citenamefont {Greenberg},
		\citenamefont {Klavins}, \citenamefont {Miranda}, \citenamefont {Gozar} \emph
		{et~al.}}]{nag2025}%
	\BibitemOpen
	\bibfield  {author} {\bibinfo {author} {\bibfnamefont {P.~K.}\ \bibnamefont
			{Nag}}, \bibinfo {author} {\bibfnamefont {K.}~\bibnamefont {Scott}}, \bibinfo
		{author} {\bibfnamefont {V.~S.}\ \bibnamefont {de~Carvalho}}, \bibinfo
		{author} {\bibfnamefont {J.~K.}\ \bibnamefont {Byland}}, \bibinfo {author}
		{\bibfnamefont {X.}~\bibnamefont {Yang}}, \bibinfo {author} {\bibfnamefont
			{M.}~\bibnamefont {Walker}}, \bibinfo {author} {\bibfnamefont {A.~G.}\
			\bibnamefont {Greenberg}}, \bibinfo {author} {\bibfnamefont {P.}~\bibnamefont
			{Klavins}}, \bibinfo {author} {\bibfnamefont {E.}~\bibnamefont {Miranda}},
		\bibinfo {author} {\bibfnamefont {A.}~\bibnamefont {Gozar}}, \emph {et~al.},\
	}\bibfield  {title} {\bibinfo {title} {Highly anisotropic superconducting gap
			near the nematic quantum critical point of
			$\mathrm{FeSe}_{1-x}\mathrm{S}_x$},\ }\href
	{https://doi.org/10.1038/s41567-024-02683-x} {\bibfield  {journal} {\bibinfo
			{journal} {Nature Physics}\ }\textbf {\bibinfo {volume} {21}},\ \bibinfo
		{pages} {89} (\bibinfo {year} {2025})}\BibitemShut {NoStop}%
	\bibitem [{\citenamefont {Binnig}\ \emph {et~al.}(1980)\citenamefont {Binnig},
		\citenamefont {Baratoff}, \citenamefont {Hoenig},\ and\ \citenamefont
		{Bednorz}}]{Binning1980}%
	\BibitemOpen
	\bibfield  {author} {\bibinfo {author} {\bibfnamefont {G.}~\bibnamefont
			{Binnig}}, \bibinfo {author} {\bibfnamefont {A.}~\bibnamefont {Baratoff}},
		\bibinfo {author} {\bibfnamefont {H.~E.}\ \bibnamefont {Hoenig}},\ and\
		\bibinfo {author} {\bibfnamefont {J.~G.}\ \bibnamefont {Bednorz}},\
	}\bibfield  {title} {\bibinfo {title} {Two-band superconductivity in
			{Nb}-doped $\mathrm{SrTiO_3}$},\ }\href
	{https://doi.org/10.1103/PhysRevLett.45.1352} {\bibfield  {journal} {\bibinfo
			{journal} {Phys. Rev. Lett.}\ }\textbf {\bibinfo {volume} {45}},\ \bibinfo
		{pages} {1352} (\bibinfo {year} {1980})}\BibitemShut {NoStop}%
	\bibitem [{\citenamefont {Lin}\ \emph {et~al.}(2014)\citenamefont {Lin},
		\citenamefont {Gourgout}, \citenamefont {Bridoux}, \citenamefont {Jomard},
		\citenamefont {Pourret}, \citenamefont {Fauqu\'e}, \citenamefont {Aoki},\
		and\ \citenamefont {Behnia}}]{Lin2014}%
	\BibitemOpen
	\bibfield  {author} {\bibinfo {author} {\bibfnamefont {X.}~\bibnamefont
			{Lin}}, \bibinfo {author} {\bibfnamefont {A.}~\bibnamefont {Gourgout}},
		\bibinfo {author} {\bibfnamefont {G.}~\bibnamefont {Bridoux}}, \bibinfo
		{author} {\bibfnamefont {F.~m.~c.}\ \bibnamefont {Jomard}}, \bibinfo {author}
		{\bibfnamefont {A.}~\bibnamefont {Pourret}}, \bibinfo {author} {\bibfnamefont
			{B.}~\bibnamefont {Fauqu\'e}}, \bibinfo {author} {\bibfnamefont
			{D.}~\bibnamefont {Aoki}},\ and\ \bibinfo {author} {\bibfnamefont
			{K.}~\bibnamefont {Behnia}},\ }\bibfield  {title} {\bibinfo {title} {Multiple
			nodeless superconducting gaps in optimally doped
			$\mathrm{SrTi}_{1\ensuremath{-}x}\mathrm{Nb}_{x}\mathrm{O}_{3}$},\ }\href
	{https://doi.org/10.1103/PhysRevB.90.140508} {\bibfield  {journal} {\bibinfo
			{journal} {Phys. Rev. B}\ }\textbf {\bibinfo {volume} {90}},\ \bibinfo
		{pages} {140508(R)} (\bibinfo {year} {2014})}\BibitemShut {NoStop}%
	\bibitem [{\citenamefont {Thiemann}\ \emph {et~al.}(2018)\citenamefont
		{Thiemann}, \citenamefont {Beutel}, \citenamefont {Dressel}, \citenamefont
		{Lee-Hone}, \citenamefont {Broun}, \citenamefont {Fillis-Tsirakis},
		\citenamefont {Boschker}, \citenamefont {Mannhart},\ and\ \citenamefont
		{Scheffler}}]{Thieman2018}%
	\BibitemOpen
	\bibfield  {author} {\bibinfo {author} {\bibfnamefont {M.}~\bibnamefont
			{Thiemann}}, \bibinfo {author} {\bibfnamefont {M.~H.}\ \bibnamefont
			{Beutel}}, \bibinfo {author} {\bibfnamefont {M.}~\bibnamefont {Dressel}},
		\bibinfo {author} {\bibfnamefont {N.~R.}\ \bibnamefont {Lee-Hone}}, \bibinfo
		{author} {\bibfnamefont {D.~M.}\ \bibnamefont {Broun}}, \bibinfo {author}
		{\bibfnamefont {E.}~\bibnamefont {Fillis-Tsirakis}}, \bibinfo {author}
		{\bibfnamefont {H.}~\bibnamefont {Boschker}}, \bibinfo {author}
		{\bibfnamefont {J.}~\bibnamefont {Mannhart}},\ and\ \bibinfo {author}
		{\bibfnamefont {M.}~\bibnamefont {Scheffler}},\ }\bibfield  {title} {\bibinfo
		{title} {Single-gap superconductivity and dome of superfluid density in
			$\mathrm{Nb}$\text{-}doped $\mathrm{SrTiO_3}$},\ }\href
	{https://doi.org/10.1103/PhysRevLett.120.237002} {\bibfield  {journal}
		{\bibinfo  {journal} {Phys. Rev. Lett.}\ }\textbf {\bibinfo {volume} {120}},\
		\bibinfo {pages} {237002} (\bibinfo {year} {2018})}\BibitemShut {NoStop}%
	\bibitem [{\citenamefont {Krishana}\ \emph {et~al.}(1999)\citenamefont
		{Krishana}, \citenamefont {Ong}, \citenamefont {Zhang}, \citenamefont {Xu},
		\citenamefont {Gagnon},\ and\ \citenamefont {Taillefer}}]{Krishana1999YBCO}%
	\BibitemOpen
	\bibfield  {author} {\bibinfo {author} {\bibfnamefont {K.}~\bibnamefont
			{Krishana}}, \bibinfo {author} {\bibfnamefont {N.~P.}\ \bibnamefont {Ong}},
		\bibinfo {author} {\bibfnamefont {Y.}~\bibnamefont {Zhang}}, \bibinfo
		{author} {\bibfnamefont {Z.~A.}\ \bibnamefont {Xu}}, \bibinfo {author}
		{\bibfnamefont {R.}~\bibnamefont {Gagnon}},\ and\ \bibinfo {author}
		{\bibfnamefont {L.}~\bibnamefont {Taillefer}},\ }\bibfield  {title} {\bibinfo
		{title} {Quasiparticle thermal hall angle and magnetoconductance in
			${{\mathrm{YBa}}_{2}{\mathrm{Cu}}_{3}\mathrm{O}}_{\mathit{x}}$},\ }\href
	{https://doi.org/10.1103/PhysRevLett.82.5108} {\bibfield  {journal} {\bibinfo
			{journal} {Phys. Rev. Lett.}\ }\textbf {\bibinfo {volume} {82}},\ \bibinfo
		{pages} {5108} (\bibinfo {year} {1999})}\BibitemShut {NoStop}%
	\bibitem [{\citenamefont {Cvetkovic}\ and\ \citenamefont
		{Vafek}(2015)}]{Cvetkovic2015}%
	\BibitemOpen
	\bibfield  {author} {\bibinfo {author} {\bibfnamefont {V.}~\bibnamefont
			{Cvetkovic}}\ and\ \bibinfo {author} {\bibfnamefont {O.}~\bibnamefont
			{Vafek}},\ }\bibfield  {title} {\bibinfo {title} {Berry phases and the
			intrinsic thermal hall effect in high-temperature cuprate superconductors},\
	}\href {https://doi.org/10.1038/ncomms7518} {\bibfield  {journal} {\bibinfo
			{journal} {Nature Communications}\ }\textbf {\bibinfo {volume} {6}},\
		\bibinfo {pages} {6518} (\bibinfo {year} {2015})}\BibitemShut {NoStop}%
	\bibitem [{\citenamefont {Altangerel}\ \emph {et~al.}(2025)\citenamefont
		{Altangerel}, \citenamefont {Barth{\'e}lemy}, \citenamefont
		{Lefran{\c{c}}ois}, \citenamefont {Baglo}, \citenamefont {Mezidi},
		\citenamefont {Grissonnanche}, \citenamefont {Vallipuram}, \citenamefont
		{Campillo}, \citenamefont {Forget}, \citenamefont {Colson} \emph
		{et~al.}}]{altangerel2025}%
	\BibitemOpen
	\bibfield  {author} {\bibinfo {author} {\bibfnamefont {M.}~\bibnamefont
			{Altangerel}}, \bibinfo {author} {\bibfnamefont {Q.}~\bibnamefont
			{Barth{\'e}lemy}}, \bibinfo {author} {\bibfnamefont {{\'E}.}~\bibnamefont
			{Lefran{\c{c}}ois}}, \bibinfo {author} {\bibfnamefont {J.}~\bibnamefont
			{Baglo}}, \bibinfo {author} {\bibfnamefont {M.}~\bibnamefont {Mezidi}},
		\bibinfo {author} {\bibfnamefont {G.}~\bibnamefont {Grissonnanche}}, \bibinfo
		{author} {\bibfnamefont {A.}~\bibnamefont {Vallipuram}}, \bibinfo {author}
		{\bibfnamefont {E.}~\bibnamefont {Campillo}}, \bibinfo {author}
		{\bibfnamefont {A.}~\bibnamefont {Forget}}, \bibinfo {author} {\bibfnamefont
			{D.}~\bibnamefont {Colson}}, \emph {et~al.},\ }\bibfield  {title} {\bibinfo
		{title} {Thermal hall conductivity in the strongest cuprate superconductor:
			Estimate of the mean free path in the trilayer cuprate
			$\mathrm{HgBa_2Ca_2Cu_3O_{8+\delta}}$},\ }\href
	{https://doi.org/10.1103/1gp5-h875} {\bibfield  {journal} {\bibinfo
			{journal} {Phys. Rev. B}\ }\textbf {\bibinfo {volume} {112}},\ \bibinfo
		{pages} {014522} (\bibinfo {year} {2025})}\BibitemShut {NoStop}%
	\bibitem [{\citenamefont {Campillo}\ \emph {et~al.}(2026)\citenamefont
		{Campillo}, \citenamefont {Mezidi}, \citenamefont {Chen}, \citenamefont
		{Vallipuram}, \citenamefont {Baglo}, \citenamefont {Altangerel},
		\citenamefont {Grissonnanche}, \citenamefont {Gu},\ and\ \citenamefont
		{Taillefer}}]{Campillo2026}%
	\BibitemOpen
	\bibfield  {author} {\bibinfo {author} {\bibfnamefont {E.}~\bibnamefont
			{Campillo}}, \bibinfo {author} {\bibfnamefont {M.}~\bibnamefont {Mezidi}},
		\bibinfo {author} {\bibfnamefont {L.}~\bibnamefont {Chen}}, \bibinfo {author}
		{\bibfnamefont {A.}~\bibnamefont {Vallipuram}}, \bibinfo {author}
		{\bibfnamefont {J.}~\bibnamefont {Baglo}}, \bibinfo {author} {\bibfnamefont
			{M.}~\bibnamefont {Altangerel}}, \bibinfo {author} {\bibfnamefont
			{G.}~\bibnamefont {Grissonnanche}}, \bibinfo {author} {\bibfnamefont
			{G.}~\bibnamefont {Gu}},\ and\ \bibinfo {author} {\bibfnamefont
			{L.}~\bibnamefont {Taillefer}},\ }\bibfield  {title} {\bibinfo {title}
		{Electronic mean free path of the cuprate superconductor
			$\mathrm{Bi_2Sr_2CaCu_2O_{8+\delta}}$ from thermal hall conductivity},\
	}\href {https://doi.org/10.1103/c4gc-hkfn} {\bibfield  {journal} {\bibinfo
			{journal} {Phys. Rev. B}\ } (\bibinfo {year} {2026})}\BibitemShut {NoStop}%
	\bibitem [{\citenamefont {Zhang}\ \emph {et~al.}(2001)\citenamefont {Zhang},
		\citenamefont {Ong}, \citenamefont {Anderson}, \citenamefont {Bonn},
		\citenamefont {Liang},\ and\ \citenamefont {Hardy}}]{Zhang2001}%
	\BibitemOpen
	\bibfield  {author} {\bibinfo {author} {\bibfnamefont {Y.}~\bibnamefont
			{Zhang}}, \bibinfo {author} {\bibfnamefont {N.~P.}\ \bibnamefont {Ong}},
		\bibinfo {author} {\bibfnamefont {P.~W.}\ \bibnamefont {Anderson}}, \bibinfo
		{author} {\bibfnamefont {D.~A.}\ \bibnamefont {Bonn}}, \bibinfo {author}
		{\bibfnamefont {R.}~\bibnamefont {Liang}},\ and\ \bibinfo {author}
		{\bibfnamefont {W.~N.}\ \bibnamefont {Hardy}},\ }\bibfield  {title} {\bibinfo
		{title} {Giant enhancement of the thermal {Hall} conductivity
			${\ensuremath{\kappa}}_{xy}$ in the superconductor
			$\mathrm{YBa}_{2}\mathrm{Cu}_{3}\mathrm{O}_{7}$},\ }\href
	{https://doi.org/10.1103/PhysRevLett.86.890} {\bibfield  {journal} {\bibinfo
			{journal} {Phys. Rev. Lett.}\ }\textbf {\bibinfo {volume} {86}},\ \bibinfo
		{pages} {890} (\bibinfo {year} {2001})}\BibitemShut {NoStop}%
	\bibitem [{\citenamefont {Kasahara}\ \emph {et~al.}(2005)\citenamefont
		{Kasahara}, \citenamefont {Nakajima}, \citenamefont {Izawa}, \citenamefont
		{Matsuda}, \citenamefont {Behnia}, \citenamefont {Shishido}, \citenamefont
		{Settai},\ and\ \citenamefont {Onuki}}]{Kasahara2005}%
	\BibitemOpen
	\bibfield  {author} {\bibinfo {author} {\bibfnamefont {Y.}~\bibnamefont
			{Kasahara}}, \bibinfo {author} {\bibfnamefont {Y.}~\bibnamefont {Nakajima}},
		\bibinfo {author} {\bibfnamefont {K.}~\bibnamefont {Izawa}}, \bibinfo
		{author} {\bibfnamefont {Y.}~\bibnamefont {Matsuda}}, \bibinfo {author}
		{\bibfnamefont {K.}~\bibnamefont {Behnia}}, \bibinfo {author} {\bibfnamefont
			{H.}~\bibnamefont {Shishido}}, \bibinfo {author} {\bibfnamefont
			{R.}~\bibnamefont {Settai}},\ and\ \bibinfo {author} {\bibfnamefont
			{Y.}~\bibnamefont {Onuki}},\ }\bibfield  {title} {\bibinfo {title} {Anomalous
			quasiparticle transport in the superconducting state of
			$\mathrm{CeCoIn}_{5}$},\ }\href {https://doi.org/10.1103/PhysRevB.72.214515}
	{\bibfield  {journal} {\bibinfo  {journal} {Phys. Rev. B}\ }\textbf {\bibinfo
			{volume} {72}},\ \bibinfo {pages} {214515} (\bibinfo {year}
		{2005})}\BibitemShut {NoStop}%
	\bibitem [{\citenamefont {Checkelsky}\ \emph {et~al.}(2012)\citenamefont
		{Checkelsky}, \citenamefont {Thomale}, \citenamefont {Li}, \citenamefont
		{Chen}, \citenamefont {Luo}, \citenamefont {Wang},\ and\ \citenamefont
		{Ong}}]{checkelsky2012thermal}%
	\BibitemOpen
	\bibfield  {author} {\bibinfo {author} {\bibfnamefont {J.}~\bibnamefont
			{Checkelsky}}, \bibinfo {author} {\bibfnamefont {R.}~\bibnamefont {Thomale}},
		\bibinfo {author} {\bibfnamefont {L.}~\bibnamefont {Li}}, \bibinfo {author}
		{\bibfnamefont {G.}~\bibnamefont {Chen}}, \bibinfo {author} {\bibfnamefont
			{J.}~\bibnamefont {Luo}}, \bibinfo {author} {\bibfnamefont {N.}~\bibnamefont
			{Wang}},\ and\ \bibinfo {author} {\bibfnamefont {N.~P.}\ \bibnamefont
			{Ong}},\ }\bibfield  {title} {\bibinfo {title} {Thermal {Hall} conductivity
			as a probe of gap structure in multiband superconductors: The case of
			$\mathrm{Ba}_{1-x}\mathrm{K}_x \mathrm{Fe_2 As_2}$},\ }\href
	{https://doi.org/10.1103/PhysRevB.86.180502} {\bibfield  {journal} {\bibinfo
			{journal} {Phys. Rev. B}\ }\textbf {\bibinfo {volume} {86}},\ \bibinfo
		{pages} {180502} (\bibinfo {year} {2012})}\BibitemShut {NoStop}%
	\bibitem [{\citenamefont {Mattheiss}(1970)}]{Mattheiss1970}%
	\BibitemOpen
	\bibfield  {author} {\bibinfo {author} {\bibfnamefont {L.~F.}\ \bibnamefont
			{Mattheiss}},\ }\bibfield  {title} {\bibinfo {title} {Electronic structure of
			niobium and tantalum},\ }\href {https://doi.org/10.1103/PhysRevB.1.373}
	{\bibfield  {journal} {\bibinfo  {journal} {Phys. Rev. B}\ }\textbf {\bibinfo
			{volume} {1}},\ \bibinfo {pages} {373} (\bibinfo {year} {1970})}\BibitemShut
	{NoStop}%
	\bibitem [{\citenamefont {Shen}\ \emph {et~al.}(1965)\citenamefont {Shen},
		\citenamefont {Senozan},\ and\ \citenamefont {Phillips}}]{shen1965evidence}%
	\BibitemOpen
	\bibfield  {author} {\bibinfo {author} {\bibfnamefont {L.~Y.~L.}\
			\bibnamefont {Shen}}, \bibinfo {author} {\bibfnamefont {N.~M.}\ \bibnamefont
			{Senozan}},\ and\ \bibinfo {author} {\bibfnamefont {N.~E.}\ \bibnamefont
			{Phillips}},\ }\bibfield  {title} {\bibinfo {title} {Evidence for two energy
			gaps in high-purity superconducting {Nb, Ta, and V}},\ }\href
	{https://doi.org/10.1103/PhysRevLett.14.1025} {\bibfield  {journal} {\bibinfo
			{journal} {Phys. Rev. Lett.}\ }\textbf {\bibinfo {volume} {14}},\ \bibinfo
		{pages} {1025} (\bibinfo {year} {1965})}\BibitemShut {NoStop}%
	\bibitem [{\citenamefont {Sung}\ and\ \citenamefont {{Yun Lung
				Shen}}(1965)}]{SUNG1965101}%
	\BibitemOpen
	\bibfield  {author} {\bibinfo {author} {\bibfnamefont {C.}~\bibnamefont
			{Sung}}\ and\ \bibinfo {author} {\bibfnamefont {L.}~\bibnamefont {{Yun Lung
					Shen}}},\ }\bibfield  {title} {\bibinfo {title} {The specific heat of
			superconducting transition metals},\ }\href
	{https://doi.org/https://doi.org/10.1016/0031-9163(65)90730-4} {\bibfield
		{journal} {\bibinfo  {journal} {Physics Letters}\ }\textbf {\bibinfo {volume}
			{19}},\ \bibinfo {pages} {101} (\bibinfo {year} {1965})}\BibitemShut
	{NoStop}%
	\bibitem [{\citenamefont {Carlson}\ and\ \citenamefont
		{Satterthwaite}(1970)}]{carlson1970anomalous}%
	\BibitemOpen
	\bibfield  {author} {\bibinfo {author} {\bibfnamefont {J.~R.}\ \bibnamefont
			{Carlson}}\ and\ \bibinfo {author} {\bibfnamefont {C.~B.}\ \bibnamefont
			{Satterthwaite}},\ }\bibfield  {title} {\bibinfo {title} {Anomalous thermal
			conductivity in superconducting niobium},\ }\href
	{https://doi.org/10.1103/PhysRevLett.24.461} {\bibfield  {journal} {\bibinfo
			{journal} {Phys. Rev. Lett.}\ }\textbf {\bibinfo {volume} {24}},\ \bibinfo
		{pages} {461} (\bibinfo {year} {1970})}\BibitemShut {NoStop}%
	\bibitem [{\citenamefont {Hafstrom}\ and\ \citenamefont
		{MacVicar}(1970)}]{hafstrom1970case}%
	\BibitemOpen
	\bibfield  {author} {\bibinfo {author} {\bibfnamefont {J.~W.}\ \bibnamefont
			{Hafstrom}}\ and\ \bibinfo {author} {\bibfnamefont {M.~L.~A.}\ \bibnamefont
			{MacVicar}},\ }\bibfield  {title} {\bibinfo {title} {Case for a second energy
			gap in superconducting niobium},\ }\href
	{https://doi.org/10.1103/PhysRevB.2.4511} {\bibfield  {journal} {\bibinfo
			{journal} {Phys. Rev. B}\ }\textbf {\bibinfo {volume} {2}},\ \bibinfo {pages}
		{4511} (\bibinfo {year} {1970})}\BibitemShut {NoStop}%
	\bibitem [{\citenamefont {Novotny}\ and\ \citenamefont
		{Meincke}(1975)}]{novotny1975single}%
	\BibitemOpen
	\bibfield  {author} {\bibinfo {author} {\bibfnamefont {V.}~\bibnamefont
			{Novotny}}\ and\ \bibinfo {author} {\bibfnamefont {P.}~\bibnamefont
			{Meincke}},\ }\bibfield  {title} {\bibinfo {title} {Single superconducting
			energy gap in pure niobium},\ }\href {https://doi.org/10.1007/BF00116976}
	{\bibfield  {journal} {\bibinfo  {journal} {Journal of Low Temperature
				Physics}\ }\textbf {\bibinfo {volume} {18}},\ \bibinfo {pages} {147}
		(\bibinfo {year} {1975})}\BibitemShut {NoStop}%
	\bibitem [{\citenamefont {Sellers}\ \emph {et~al.}(1973)\citenamefont
		{Sellers}, \citenamefont {Anderson},\ and\ \citenamefont
		{Birnbaum}}]{sellers1973anomalous}%
	\BibitemOpen
	\bibfield  {author} {\bibinfo {author} {\bibfnamefont {G.}~\bibnamefont
			{Sellers}}, \bibinfo {author} {\bibfnamefont {A.}~\bibnamefont {Anderson}},\
		and\ \bibinfo {author} {\bibfnamefont {H.}~\bibnamefont {Birnbaum}},\
	}\bibfield  {title} {\bibinfo {title} {The anomalous heat capacity of
			superconducting niobium},\ }\href
	{https://doi.org/10.1016/0375-9601(73)90870-0} {\bibfield  {journal}
		{\bibinfo  {journal} {Physics Letters A}\ }\textbf {\bibinfo {volume} {44}},\
		\bibinfo {pages} {173} (\bibinfo {year} {1973})}\BibitemShut {NoStop}%
	\bibitem [{\citenamefont {Anderson}\ \emph {et~al.}(1971)\citenamefont
		{Anderson}, \citenamefont {Satterthwaite},\ and\ \citenamefont
		{Smith}}]{Anderson1971}%
	\BibitemOpen
	\bibfield  {author} {\bibinfo {author} {\bibfnamefont {A.~C.}\ \bibnamefont
			{Anderson}}, \bibinfo {author} {\bibfnamefont {C.~B.}\ \bibnamefont
			{Satterthwaite}},\ and\ \bibinfo {author} {\bibfnamefont {S.~C.}\
			\bibnamefont {Smith}},\ }\bibfield  {title} {\bibinfo {title} {Thermal
			conductivity of superconducting niobium},\ }\href
	{https://doi.org/10.1103/PhysRevB.3.3762} {\bibfield  {journal} {\bibinfo
			{journal} {Phys. Rev. B}\ }\textbf {\bibinfo {volume} {3}},\ \bibinfo {pages}
		{3762} (\bibinfo {year} {1971})}\BibitemShut {NoStop}%
	\bibitem [{\citenamefont {Almond}\ \emph {et~al.}(1972)\citenamefont {Almond},
		\citenamefont {Lea},\ and\ \citenamefont {Dobbs}}]{Almond1972}%
	\BibitemOpen
	\bibfield  {author} {\bibinfo {author} {\bibfnamefont {D.~P.}\ \bibnamefont
			{Almond}}, \bibinfo {author} {\bibfnamefont {M.~J.}\ \bibnamefont {Lea}},\
		and\ \bibinfo {author} {\bibfnamefont {E.~R.}\ \bibnamefont {Dobbs}},\
	}\bibfield  {title} {\bibinfo {title} {Ultrasonic evidence against multiple
			energy gaps in superconducting niobium},\ }\href
	{https://doi.org/10.1103/PhysRevLett.29.764} {\bibfield  {journal} {\bibinfo
			{journal} {Phys. Rev. Lett.}\ }\textbf {\bibinfo {volume} {29}},\ \bibinfo
		{pages} {764} (\bibinfo {year} {1972})}\BibitemShut {NoStop}%
	\bibitem [{\citenamefont {Noto}(1969)}]{noto1969}%
	\BibitemOpen
	\bibfield  {author} {\bibinfo {author} {\bibfnamefont {K.}~\bibnamefont
			{Noto}},\ }\bibfield  {title} {\bibinfo {title} {Thermal conductivity in a
			pure type {II} superconductor near the upper critical field},\ }\href
	{https://doi.org/10.1143/JPSJ.26.710} {\bibfield  {journal} {\bibinfo
			{journal} {Journal of the Physical Society of Japan}\ }\textbf {\bibinfo
			{volume} {26}},\ \bibinfo {pages} {710} (\bibinfo {year} {1969})}\BibitemShut
	{NoStop}%
	\bibitem [{\citenamefont {Lowell}\ and\ \citenamefont
		{Sousa}(1970)}]{Lowell1970}%
	\BibitemOpen
	\bibfield  {author} {\bibinfo {author} {\bibfnamefont {J.}~\bibnamefont
			{Lowell}}\ and\ \bibinfo {author} {\bibfnamefont {J.~B.}\ \bibnamefont
			{Sousa}},\ }\bibfield  {title} {\bibinfo {title} {Mixed-state thermal
			conductivity of type {II} superconductors},\ }\href
	{https://doi.org/10.1007/BF00628399} {\bibfield  {journal} {\bibinfo
			{journal} {Journal of Low Temperature Physics}\ }\textbf {\bibinfo {volume}
			{3}},\ \bibinfo {pages} {65} (\bibinfo {year} {1970})}\BibitemShut {NoStop}%
	\bibitem [{\citenamefont {Luo}\ and\ \citenamefont {Zhang}(2025)}]{Luo2025}%
	\BibitemOpen
	\bibfield  {author} {\bibinfo {author} {\bibfnamefont {D.}~\bibnamefont
			{Luo}}\ and\ \bibinfo {author} {\bibfnamefont {J.}~\bibnamefont {Zhang}},\
	}\bibfield  {title} {\bibinfo {title} {The magnetic field impact on the
			thermal conductivity of superconducting niobium},\ }\href
	{https://doi.org/10.1088/1742-6596/2953/1/012005} {\bibfield  {journal}
		{\bibinfo  {journal} {Journal of Physics: Conference Series}\ }\textbf
		{\bibinfo {volume} {2953}},\ \bibinfo {pages} {012005} (\bibinfo {year}
		{2025})}\BibitemShut {NoStop}%
	\bibitem [{\citenamefont {Shakeripour}\ \emph {et~al.}(2009)\citenamefont
		{Shakeripour}, \citenamefont {Petrovic},\ and\ \citenamefont
		{Taillefer}}]{Shakeripour2009}%
	\BibitemOpen
	\bibfield  {author} {\bibinfo {author} {\bibfnamefont {H.}~\bibnamefont
			{Shakeripour}}, \bibinfo {author} {\bibfnamefont {C.}~\bibnamefont
			{Petrovic}},\ and\ \bibinfo {author} {\bibfnamefont {L.}~\bibnamefont
			{Taillefer}},\ }\bibfield  {title} {\bibinfo {title} {Heat transport as a
			probe of superconducting gap structure},\ }\href
	{https://doi.org/10.1088/1367-2630/11/5/055065} {\bibfield  {journal}
		{\bibinfo  {journal} {New Journal of Physics}\ }\textbf {\bibinfo {volume}
			{11}},\ \bibinfo {pages} {055065} (\bibinfo {year} {2009})}\BibitemShut
	{NoStop}%
	\bibitem [{\citenamefont {Bardeen}\ \emph {et~al.}(1959)\citenamefont
		{Bardeen}, \citenamefont {Rickayzen},\ and\ \citenamefont
		{Tewordt}}]{Bardeen1959}%
	\BibitemOpen
	\bibfield  {author} {\bibinfo {author} {\bibfnamefont {J.}~\bibnamefont
			{Bardeen}}, \bibinfo {author} {\bibfnamefont {G.}~\bibnamefont {Rickayzen}},\
		and\ \bibinfo {author} {\bibfnamefont {L.}~\bibnamefont {Tewordt}},\
	}\bibfield  {title} {\bibinfo {title} {Theory of the thermal conductivity of
			superconductors},\ }\href {https://doi.org/10.1103/PhysRev.113.982}
	{\bibfield  {journal} {\bibinfo  {journal} {Phys. Rev.}\ }\textbf {\bibinfo
			{volume} {113}},\ \bibinfo {pages} {982} (\bibinfo {year}
		{1959})}\BibitemShut {NoStop}%
	\bibitem [{\citenamefont {Fawcett}\ \emph {et~al.}(1967)\citenamefont
		{Fawcett}, \citenamefont {Reed},\ and\ \citenamefont {Soden}}]{Fawcett1967}%
	\BibitemOpen
	\bibfield  {author} {\bibinfo {author} {\bibfnamefont {E.}~\bibnamefont
			{Fawcett}}, \bibinfo {author} {\bibfnamefont {W.~A.}\ \bibnamefont {Reed}},\
		and\ \bibinfo {author} {\bibfnamefont {R.~R.}\ \bibnamefont {Soden}},\
	}\bibfield  {title} {\bibinfo {title} {High-field galvanomagnetic properties
			of niobium and tantalum},\ }\href {https://doi.org/10.1103/PhysRev.159.533}
	{\bibfield  {journal} {\bibinfo  {journal} {Physical Review}\ }\textbf
		{\bibinfo {volume} {159}},\ \bibinfo {pages} {533} (\bibinfo {year}
		{1967})}\BibitemShut {NoStop}%
	\bibitem [{\citenamefont {Kes}\ \emph {et~al.}(1975)\citenamefont {Kes},
		\citenamefont {Van~der Veeken},\ and\ \citenamefont
		{De~Klerk}}]{kes1975thermal}%
	\BibitemOpen
	\bibfield  {author} {\bibinfo {author} {\bibfnamefont {P.}~\bibnamefont
			{Kes}}, \bibinfo {author} {\bibfnamefont {J.}~\bibnamefont {Van~der
				Veeken}},\ and\ \bibinfo {author} {\bibfnamefont {D.}~\bibnamefont
			{De~Klerk}},\ }\bibfield  {title} {\bibinfo {title} {Thermal conductivity of
			niobium in the mixed state},\ }\href {https://doi.org/10.1007/BF00118165}
	{\bibfield  {journal} {\bibinfo  {journal} {Journal of Low Temperature
				Physics}\ }\textbf {\bibinfo {volume} {18}},\ \bibinfo {pages} {355}
		(\bibinfo {year} {1975})}\BibitemShut {NoStop}%
	\bibitem [{\citenamefont {Chakal'ski}\ \emph {et~al.}(1978)\citenamefont
		{Chakal'ski}, \citenamefont {Red'ko}, \citenamefont {Shalyt},\ and\
		\citenamefont {M.}}]{Chakalskii1978}%
	\BibitemOpen
	\bibfield  {author} {\bibinfo {author} {\bibfnamefont {B.}~\bibnamefont
			{Chakal'ski}}, \bibinfo {author} {\bibfnamefont {N.~A.}\ \bibnamefont
			{Red'ko}}, \bibinfo {author} {\bibfnamefont {S.~S.}\ \bibnamefont {Shalyt}},\
		and\ \bibinfo {author} {\bibfnamefont {A.~V.}\ \bibnamefont {M.}},\
	}\bibfield  {title} {\bibinfo {title} {Thermal conductivity of pure vanadium
			in normal, superconducting, and mixed states},\ }\href
	{https://www.osti.gov/etdeweb/biblio/5680683} {\bibfield  {journal} {\bibinfo
			{journal} {Journal of Experimental and Theoretical Physics}\ }\textbf
		{\bibinfo {volume} {48}},\ \bibinfo {pages} {665} (\bibinfo {year}
		{1978})}\BibitemShut {NoStop}%
	\bibitem [{\citenamefont {Behnia}\ \emph {et~al.}(1991)\citenamefont {Behnia},
		\citenamefont {Taillefer}, \citenamefont {Flouquet}, \citenamefont {Jaccard},
		\citenamefont {Maki},\ and\ \citenamefont {Fisk}}]{Behnia1991}%
	\BibitemOpen
	\bibfield  {author} {\bibinfo {author} {\bibfnamefont {K.}~\bibnamefont
			{Behnia}}, \bibinfo {author} {\bibfnamefont {L.}~\bibnamefont {Taillefer}},
		\bibinfo {author} {\bibfnamefont {J.}~\bibnamefont {Flouquet}}, \bibinfo
		{author} {\bibfnamefont {D.}~\bibnamefont {Jaccard}}, \bibinfo {author}
		{\bibfnamefont {K.}~\bibnamefont {Maki}},\ and\ \bibinfo {author}
		{\bibfnamefont {Z.}~\bibnamefont {Fisk}},\ }\bibfield  {title} {\bibinfo
		{title} {Thermal conductivity of superconducting $\mathrm{UPt}_3$},\ }\href
	{https://doi.org/10.1007/BF00683610} {\bibfield  {journal} {\bibinfo
			{journal} {Journal of Low Temperature Physics}\ }\textbf {\bibinfo {volume}
			{84}},\ \bibinfo {pages} {261} (\bibinfo {year} {1991})}\BibitemShut
	{NoStop}%
	\bibitem [{\citenamefont {Belin}\ \emph {et~al.}(1998)\citenamefont {Belin},
		\citenamefont {Behnia},\ and\ \citenamefont {Deluzet}}]{Belin1998}%
	\BibitemOpen
	\bibfield  {author} {\bibinfo {author} {\bibfnamefont {S.}~\bibnamefont
			{Belin}}, \bibinfo {author} {\bibfnamefont {K.}~\bibnamefont {Behnia}},\ and\
		\bibinfo {author} {\bibfnamefont {A.}~\bibnamefont {Deluzet}},\ }\bibfield
	{title} {\bibinfo {title} {Heat conduction in
			$\mathit{\ensuremath{\kappa}}\ensuremath{-}(\mathrm{BEDT}\ensuremath{-}\mathrm{TTF}{)}_{2}\mathrm{Cu}(\mathrm{NCS}{)}_{2}$},\
	}\href {https://doi.org/10.1103/PhysRevLett.81.4728} {\bibfield  {journal}
		{\bibinfo  {journal} {Phys. Rev. Lett.}\ }\textbf {\bibinfo {volume} {81}},\
		\bibinfo {pages} {4728} (\bibinfo {year} {1998})}\BibitemShut {NoStop}%
	\bibitem [{\citenamefont {Izawa}\ \emph {et~al.}(2001)\citenamefont {Izawa},
		\citenamefont {Yamaguchi}, \citenamefont {Sasaki},\ and\ \citenamefont
		{Matsuda}}]{Izawa2001}%
	\BibitemOpen
	\bibfield  {author} {\bibinfo {author} {\bibfnamefont {K.}~\bibnamefont
			{Izawa}}, \bibinfo {author} {\bibfnamefont {H.}~\bibnamefont {Yamaguchi}},
		\bibinfo {author} {\bibfnamefont {T.}~\bibnamefont {Sasaki}},\ and\ \bibinfo
		{author} {\bibfnamefont {Y.}~\bibnamefont {Matsuda}},\ }\bibfield  {title}
	{\bibinfo {title} {Superconducting gap structure of
			$\mathit{\ensuremath{\kappa}}\ensuremath{-}(\mathrm{BEDT}\ensuremath{-}\mathrm{TTF}{)}_{2}\mathrm{Cu}(\mathrm{NCS}{)}_{2}$
			probed by thermal conductivity tensor},\ }\href
	{https://doi.org/10.1103/PhysRevLett.88.027002} {\bibfield  {journal}
		{\bibinfo  {journal} {Phys. Rev. Lett.}\ }\textbf {\bibinfo {volume} {88}},\
		\bibinfo {pages} {027002} (\bibinfo {year} {2001})}\BibitemShut {NoStop}%
	\bibitem [{\citenamefont {Li}\ \emph {et~al.}(2020)\citenamefont {Li},
		\citenamefont {Fauqu\'e}, \citenamefont {Zhu},\ and\ \citenamefont
		{Behnia}}]{Li2020}%
	\BibitemOpen
	\bibfield  {author} {\bibinfo {author} {\bibfnamefont {X.}~\bibnamefont
			{Li}}, \bibinfo {author} {\bibfnamefont {B.}~\bibnamefont {Fauqu\'e}},
		\bibinfo {author} {\bibfnamefont {Z.}~\bibnamefont {Zhu}},\ and\ \bibinfo
		{author} {\bibfnamefont {K.}~\bibnamefont {Behnia}},\ }\bibfield  {title}
	{\bibinfo {title} {Phonon thermal {Hall} effect in strontium titanate},\
	}\href {https://doi.org/10.1103/PhysRevLett.124.105901} {\bibfield  {journal}
		{\bibinfo  {journal} {Phys. Rev. Lett.}\ }\textbf {\bibinfo {volume} {124}},\
		\bibinfo {pages} {105901} (\bibinfo {year} {2020})}\BibitemShut {NoStop}%
	\bibitem [{\citenamefont {Grissonnanche}\ \emph {et~al.}(2019)\citenamefont
		{Grissonnanche}, \citenamefont {Legros}, \citenamefont {Badoux},
		\citenamefont {Lefran{\c{c}}ois}, \citenamefont {Zatko}, \citenamefont
		{Lizaire}, \citenamefont {Lalibert{\'e}}, \citenamefont {Gourgout},
		\citenamefont {Zhou}, \citenamefont {Pyon} \emph
		{et~al.}}]{grissonnanche2019}%
	\BibitemOpen
	\bibfield  {author} {\bibinfo {author} {\bibfnamefont {G.}~\bibnamefont
			{Grissonnanche}}, \bibinfo {author} {\bibfnamefont {A.}~\bibnamefont
			{Legros}}, \bibinfo {author} {\bibfnamefont {S.}~\bibnamefont {Badoux}},
		\bibinfo {author} {\bibfnamefont {E.}~\bibnamefont {Lefran{\c{c}}ois}},
		\bibinfo {author} {\bibfnamefont {V.}~\bibnamefont {Zatko}}, \bibinfo
		{author} {\bibfnamefont {M.}~\bibnamefont {Lizaire}}, \bibinfo {author}
		{\bibfnamefont {F.}~\bibnamefont {Lalibert{\'e}}}, \bibinfo {author}
		{\bibfnamefont {A.}~\bibnamefont {Gourgout}}, \bibinfo {author}
		{\bibfnamefont {J.-S.}\ \bibnamefont {Zhou}}, \bibinfo {author}
		{\bibfnamefont {S.}~\bibnamefont {Pyon}}, \emph {et~al.},\ }\bibfield
	{title} {\bibinfo {title} {Giant thermal {Hall} conductivity in the pseudogap
			phase of cuprate superconductors},\ }\href
	{https://doi.org/10.1038/s41586-019-1375-0} {\bibfield  {journal} {\bibinfo
			{journal} {Nature}\ }\textbf {\bibinfo {volume} {571}},\ \bibinfo {pages}
		{376} (\bibinfo {year} {2019})}\BibitemShut {NoStop}%
	\bibitem [{\citenamefont {Grissonnanche}\ \emph {et~al.}(2020)\citenamefont
		{Grissonnanche}, \citenamefont {Th{\'e}riault}, \citenamefont {Gourgout},
		\citenamefont {Boulanger}, \citenamefont {Lefran{\c{c}}ois}, \citenamefont
		{Ataei}, \citenamefont {Lalibert{\'e}}, \citenamefont {Dion}, \citenamefont
		{Zhou}, \citenamefont {Pyon} \emph {et~al.}}]{grissonnanche2020chiral}%
	\BibitemOpen
	\bibfield  {author} {\bibinfo {author} {\bibfnamefont {G.}~\bibnamefont
			{Grissonnanche}}, \bibinfo {author} {\bibfnamefont {S.}~\bibnamefont
			{Th{\'e}riault}}, \bibinfo {author} {\bibfnamefont {A.}~\bibnamefont
			{Gourgout}}, \bibinfo {author} {\bibfnamefont {M.-E.}\ \bibnamefont
			{Boulanger}}, \bibinfo {author} {\bibfnamefont {E.}~\bibnamefont
			{Lefran{\c{c}}ois}}, \bibinfo {author} {\bibfnamefont {A.}~\bibnamefont
			{Ataei}}, \bibinfo {author} {\bibfnamefont {F.}~\bibnamefont
			{Lalibert{\'e}}}, \bibinfo {author} {\bibfnamefont {M.}~\bibnamefont {Dion}},
		\bibinfo {author} {\bibfnamefont {J.-S.}\ \bibnamefont {Zhou}}, \bibinfo
		{author} {\bibfnamefont {S.}~\bibnamefont {Pyon}}, \emph {et~al.},\
	}\bibfield  {title} {\bibinfo {title} {Chiral phonons in the pseudogap phase
			of cuprates},\ }\href {https://doi.org/10.1038/s41567-020-0965-y} {\bibfield
		{journal} {\bibinfo  {journal} {Nature Physics}\ }\textbf {\bibinfo {volume}
			{16}},\ \bibinfo {pages} {1108} (\bibinfo {year} {2020})}\BibitemShut
	{NoStop}%
	\bibitem [{\citenamefont {Boulanger}\ \emph {et~al.}(2020)\citenamefont
		{Boulanger}, \citenamefont {Grissonnanche}, \citenamefont {Badoux},
		\citenamefont {Allaire}, \citenamefont {Lefran{\c{c}}ois}, \citenamefont
		{Legros}, \citenamefont {Gourgout}, \citenamefont {Dion}, \citenamefont
		{Wang}, \citenamefont {Chen} \emph {et~al.}}]{boulanger2020}%
	\BibitemOpen
	\bibfield  {author} {\bibinfo {author} {\bibfnamefont {M.-E.}\ \bibnamefont
			{Boulanger}}, \bibinfo {author} {\bibfnamefont {G.}~\bibnamefont
			{Grissonnanche}}, \bibinfo {author} {\bibfnamefont {S.}~\bibnamefont
			{Badoux}}, \bibinfo {author} {\bibfnamefont {A.}~\bibnamefont {Allaire}},
		\bibinfo {author} {\bibfnamefont {{\'E}.}~\bibnamefont {Lefran{\c{c}}ois}},
		\bibinfo {author} {\bibfnamefont {A.}~\bibnamefont {Legros}}, \bibinfo
		{author} {\bibfnamefont {A.}~\bibnamefont {Gourgout}}, \bibinfo {author}
		{\bibfnamefont {M.}~\bibnamefont {Dion}}, \bibinfo {author} {\bibfnamefont
			{C.}~\bibnamefont {Wang}}, \bibinfo {author} {\bibfnamefont {X.}~\bibnamefont
			{Chen}}, \emph {et~al.},\ }\bibfield  {title} {\bibinfo {title} {Thermal
			{Hall} conductivity in the cuprate {Mott} insulators $\mathrm{Nd_2CuO_4}$ and
			$\mathrm{Sr_2CuO_2Cl_2}$},\ }\href
	{https://doi.org/10.1038/s41467-020-18881-z} {\bibfield  {journal} {\bibinfo
			{journal} {Nature communications}\ }\textbf {\bibinfo {volume} {11}},\
		\bibinfo {pages} {5325} (\bibinfo {year} {2020})}\BibitemShut {NoStop}%
	\bibitem [{\citenamefont {Li}\ \emph {et~al.}(2023)\citenamefont {Li},
		\citenamefont {Machida}, \citenamefont {Subedi}, \citenamefont {Zhu},
		\citenamefont {Li},\ and\ \citenamefont {Behnia}}]{li2023phonon}%
	\BibitemOpen
	\bibfield  {author} {\bibinfo {author} {\bibfnamefont {X.}~\bibnamefont
			{Li}}, \bibinfo {author} {\bibfnamefont {Y.}~\bibnamefont {Machida}},
		\bibinfo {author} {\bibfnamefont {A.}~\bibnamefont {Subedi}}, \bibinfo
		{author} {\bibfnamefont {Z.}~\bibnamefont {Zhu}}, \bibinfo {author}
		{\bibfnamefont {L.}~\bibnamefont {Li}},\ and\ \bibinfo {author}
		{\bibfnamefont {K.}~\bibnamefont {Behnia}},\ }\bibfield  {title} {\bibinfo
		{title} {The phonon thermal {Hall} angle in black phosphorus},\ }\href
	{https://doi.org/10.1038/s41467-023-36750-3} {\bibfield  {journal} {\bibinfo
			{journal} {Nature Communications}\ }\textbf {\bibinfo {volume} {14}},\
		\bibinfo {pages} {1027} (\bibinfo {year} {2023})}\BibitemShut {NoStop}%
	\bibitem [{\citenamefont {Jin}\ \emph {et~al.}(2025)\citenamefont {Jin},
		\citenamefont {Zhang}, \citenamefont {Wan}, \citenamefont {Wang},
		\citenamefont {Jiao},\ and\ \citenamefont {Li}}]{Jin2025}%
	\BibitemOpen
	\bibfield  {author} {\bibinfo {author} {\bibfnamefont {X.~B.}\ \bibnamefont
			{Jin}}, \bibinfo {author} {\bibfnamefont {X.}~\bibnamefont {Zhang}}, \bibinfo
		{author} {\bibfnamefont {W.~B.}\ \bibnamefont {Wan}}, \bibinfo {author}
		{\bibfnamefont {H.~R.}\ \bibnamefont {Wang}}, \bibinfo {author}
		{\bibfnamefont {Y.~H.}\ \bibnamefont {Jiao}},\ and\ \bibinfo {author}
		{\bibfnamefont {S.~Y.}\ \bibnamefont {Li}},\ }\bibfield  {title} {\bibinfo
		{title} {Discovery of universal phonon thermal {Hall} effect in crystals},\
	}\href {https://doi.org/10.1103/r572-5dfm} {\bibfield  {journal} {\bibinfo
			{journal} {Phys. Rev. Lett.}\ }\textbf {\bibinfo {volume} {135}},\ \bibinfo
		{pages} {196302} (\bibinfo {year} {2025})}\BibitemShut {NoStop}%
	\bibitem [{\citenamefont {Boulanger}\ \emph {et~al.}(2022)\citenamefont
		{Boulanger}, \citenamefont {Grissonnanche}, \citenamefont
		{Lefran\ifmmode~\mbox{\c{c}}\else \c{c}\fi{}ois}, \citenamefont {Gourgout},
		\citenamefont {Xu}, \citenamefont {Shen}, \citenamefont {Greene},\ and\
		\citenamefont {Taillefer}}]{Boulanger2022}%
	\BibitemOpen
	\bibfield  {author} {\bibinfo {author} {\bibfnamefont {M.-E.}\ \bibnamefont
			{Boulanger}}, \bibinfo {author} {\bibfnamefont {G.}~\bibnamefont
			{Grissonnanche}}, \bibinfo {author} {\bibfnamefont {E.}~\bibnamefont
			{Lefran\ifmmode~\mbox{\c{c}}\else \c{c}\fi{}ois}}, \bibinfo {author}
		{\bibfnamefont {A.}~\bibnamefont {Gourgout}}, \bibinfo {author}
		{\bibfnamefont {K.-J.}\ \bibnamefont {Xu}}, \bibinfo {author} {\bibfnamefont
			{Z.-X.}\ \bibnamefont {Shen}}, \bibinfo {author} {\bibfnamefont {R.~L.}\
			\bibnamefont {Greene}},\ and\ \bibinfo {author} {\bibfnamefont
			{L.}~\bibnamefont {Taillefer}},\ }\bibfield  {title} {\bibinfo {title}
		{Thermal {Hall} conductivity of electron-doped cuprates},\ }\href
	{https://doi.org/10.1103/PhysRevB.105.115101} {\bibfield  {journal} {\bibinfo
			{journal} {Phys. Rev. B}\ }\textbf {\bibinfo {volume} {105}},\ \bibinfo
		{pages} {115101} (\bibinfo {year} {2022})}\BibitemShut {NoStop}%
	\bibitem [{\citenamefont {Gross}\ \emph {et~al.}(1986)\citenamefont {Gross},
		\citenamefont {Chandrasekhar}, \citenamefont {Einzel}, \citenamefont
		{Andres}, \citenamefont {Hirschfeld}, \citenamefont {Ott}, \citenamefont
		{Beuers}, \citenamefont {Fisk},\ and\ \citenamefont
		{Smith}}]{gross1986anomalous}%
	\BibitemOpen
	\bibfield  {author} {\bibinfo {author} {\bibfnamefont {F.}~\bibnamefont
			{Gross}}, \bibinfo {author} {\bibfnamefont {B.}~\bibnamefont
			{Chandrasekhar}}, \bibinfo {author} {\bibfnamefont {D.}~\bibnamefont
			{Einzel}}, \bibinfo {author} {\bibfnamefont {K.}~\bibnamefont {Andres}},
		\bibinfo {author} {\bibfnamefont {P.}~\bibnamefont {Hirschfeld}}, \bibinfo
		{author} {\bibfnamefont {H.}~\bibnamefont {Ott}}, \bibinfo {author}
		{\bibfnamefont {J.}~\bibnamefont {Beuers}}, \bibinfo {author} {\bibfnamefont
			{Z.}~\bibnamefont {Fisk}},\ and\ \bibinfo {author} {\bibfnamefont
			{J.}~\bibnamefont {Smith}},\ }\bibfield  {title} {\bibinfo {title} {Anomalous
			temperature dependence of the magnetic field penetration depth in
			superconducting $\mathrm{UBe_{13}}$},\ }\href
	{https://doi.org/10.1007/BF01303700} {\bibfield  {journal} {\bibinfo
			{journal} {Zeitschrift f{\"u}r Physik B Condensed Matter}\ }\textbf {\bibinfo
			{volume} {64}},\ \bibinfo {pages} {175} (\bibinfo {year} {1986})}\BibitemShut
	{NoStop}%
	\bibitem [{\citenamefont {Hirshfeld}\ \emph {et~al.}(1962)\citenamefont
		{Hirshfeld}, \citenamefont {Leupold},\ and\ \citenamefont
		{Boorse}}]{Hirshfeld1962}%
	\BibitemOpen
	\bibfield  {author} {\bibinfo {author} {\bibfnamefont {A.~T.}\ \bibnamefont
			{Hirshfeld}}, \bibinfo {author} {\bibfnamefont {H.~A.}\ \bibnamefont
			{Leupold}},\ and\ \bibinfo {author} {\bibfnamefont {H.~A.}\ \bibnamefont
			{Boorse}},\ }\bibfield  {title} {\bibinfo {title} {Superconducting and normal
			specific heats of niobium},\ }\href
	{https://doi.org/10.1103/PhysRev.127.1501} {\bibfield  {journal} {\bibinfo
			{journal} {Phys. Rev.}\ }\textbf {\bibinfo {volume} {127}},\ \bibinfo {pages}
		{1501} (\bibinfo {year} {1962})}\BibitemShut {NoStop}%
	\bibitem [{\citenamefont {Connolly}\ and\ \citenamefont
		{Mendelssohn}(1962)}]{connolly1962}%
	\BibitemOpen
	\bibfield  {author} {\bibinfo {author} {\bibfnamefont {A.}~\bibnamefont
			{Connolly}}\ and\ \bibinfo {author} {\bibfnamefont {K.~A.~G.}\ \bibnamefont
			{Mendelssohn}},\ }\bibfield  {title} {\bibinfo {title} {Thermal conductivity
			of tantalum and niobium below 1 {K}},\ }\href
	{https://doi.org/10.1098/rspa.1962.0071} {\bibfield  {journal} {\bibinfo
			{journal} {Proceedings of the Royal Society of London. Series A. Mathematical
				and Physical Sciences}\ }\textbf {\bibinfo {volume} {266}},\ \bibinfo {pages}
		{429} (\bibinfo {year} {1962})}\BibitemShut {NoStop}%
	\bibitem [{\citenamefont {van~der Hoeven}\ and\ \citenamefont
		{Keesom}(1964)}]{van1964}%
	\BibitemOpen
	\bibfield  {author} {\bibinfo {author} {\bibfnamefont {B.~J.~C.}\
			\bibnamefont {van~der Hoeven}}\ and\ \bibinfo {author} {\bibfnamefont
			{P.~H.}\ \bibnamefont {Keesom}},\ }\bibfield  {title} {\bibinfo {title}
		{Specific heat of niobium between 0.4 and
			4.2\ifmmode^\circ\else\textdegree\fi{K}},\ }\href
	{https://doi.org/10.1103/PhysRev.134.A1320} {\bibfield  {journal} {\bibinfo
			{journal} {Phys. Rev.}\ }\textbf {\bibinfo {volume} {134}},\ \bibinfo {pages}
		{A1320} (\bibinfo {year} {1964})}\BibitemShut {NoStop}%
	\bibitem [{\citenamefont {Wasim}\ and\ \citenamefont
		{Zebouni}(1969)}]{Wasim1969}%
	\BibitemOpen
	\bibfield  {author} {\bibinfo {author} {\bibfnamefont {S.~M.}\ \bibnamefont
			{Wasim}}\ and\ \bibinfo {author} {\bibfnamefont {N.~H.}\ \bibnamefont
			{Zebouni}},\ }\bibfield  {title} {\bibinfo {title} {Thermal conductivity of
			superconducting niobium},\ }\href {https://doi.org/10.1103/PhysRev.187.539}
	{\bibfield  {journal} {\bibinfo  {journal} {Phys. Rev.}\ }\textbf {\bibinfo
			{volume} {187}},\ \bibinfo {pages} {539} (\bibinfo {year}
		{1969})}\BibitemShut {NoStop}%
	\bibitem [{\citenamefont {Kes}\ \emph {et~al.}(1974)\citenamefont {Kes},
		\citenamefont {Rolfes},\ and\ \citenamefont {De~Klerk}}]{kes1974thermal}%
	\BibitemOpen
	\bibfield  {author} {\bibinfo {author} {\bibfnamefont {P.}~\bibnamefont
			{Kes}}, \bibinfo {author} {\bibfnamefont {J.}~\bibnamefont {Rolfes}},\ and\
		\bibinfo {author} {\bibfnamefont {D.}~\bibnamefont {De~Klerk}},\ }\bibfield
	{title} {\bibinfo {title} {Thermal conductivity of niobium in the purely
			superconducting and normal states},\ }\href
	{https://doi.org/10.1007/BF00659079} {\bibfield  {journal} {\bibinfo
			{journal} {Journal of Low Temperature Physics}\ }\textbf {\bibinfo {volume}
			{17}},\ \bibinfo {pages} {341} (\bibinfo {year} {1974})}\BibitemShut
	{NoStop}%
	\bibitem [{\citenamefont {Mamyia}\ \emph {et~al.}(1974)\citenamefont {Mamyia},
		\citenamefont {Oota},\ and\ \citenamefont {Masuda}}]{mamyia1974thermal}%
	\BibitemOpen
	\bibfield  {author} {\bibinfo {author} {\bibfnamefont {T.}~\bibnamefont
			{Mamyia}}, \bibinfo {author} {\bibfnamefont {A.}~\bibnamefont {Oota}},\ and\
		\bibinfo {author} {\bibfnamefont {Y.}~\bibnamefont {Masuda}},\ }\bibfield
	{title} {\bibinfo {title} {Thermal conductivity of superconducting niobium},\
	}\href {https://doi.org/10.1016/0038-1098(74)91213-7} {\bibfield  {journal}
		{\bibinfo  {journal} {Solid State Communications}\ }\textbf {\bibinfo
			{volume} {15}},\ \bibinfo {pages} {1689} (\bibinfo {year}
		{1974})}\BibitemShut {NoStop}%
	\bibitem [{\citenamefont {Gladun}\ \emph {et~al.}(1977)\citenamefont {Gladun},
		\citenamefont {Gladun}, \citenamefont {Knorn},\ and\ \citenamefont
		{Vinzelberg}}]{gladun1977}%
	\BibitemOpen
	\bibfield  {author} {\bibinfo {author} {\bibfnamefont {A.}~\bibnamefont
			{Gladun}}, \bibinfo {author} {\bibfnamefont {C.}~\bibnamefont {Gladun}},
		\bibinfo {author} {\bibfnamefont {M.}~\bibnamefont {Knorn}},\ and\ \bibinfo
		{author} {\bibfnamefont {H.}~\bibnamefont {Vinzelberg}},\ }\bibfield  {title}
	{\bibinfo {title} {Investigation of the heat conductivity of niobium in the
			temperature range 0.05--23 {K}},\ }\href {https://doi.org/10.1007/BF00655712}
	{\bibfield  {journal} {\bibinfo  {journal} {Journal of Low Temperature
				Physics}\ }\textbf {\bibinfo {volume} {27}},\ \bibinfo {pages} {873}
		(\bibinfo {year} {1977})}\BibitemShut {NoStop}%
	\bibitem [{\citenamefont {Caroli}\ and\ \citenamefont
		{Cyrot}(1965)}]{caroli1965}%
	\BibitemOpen
	\bibfield  {author} {\bibinfo {author} {\bibfnamefont {C.}~\bibnamefont
			{Caroli}}\ and\ \bibinfo {author} {\bibfnamefont {M.}~\bibnamefont {Cyrot}},\
	}\bibfield  {title} {\bibinfo {title} {Thermal conductivity in dirty
			superconducting alloys in high field},\ }\href
	{https://doi.org/10.1007/BF02422842} {\bibfield  {journal} {\bibinfo
			{journal} {Physik der kondensierten Materie}\ }\textbf {\bibinfo {volume}
			{4}},\ \bibinfo {pages} {285} (\bibinfo {year} {1965})}\BibitemShut {NoStop}%
	\bibitem [{\citenamefont {Maki}(1967)}]{Maki1967}%
	\BibitemOpen
	\bibfield  {author} {\bibinfo {author} {\bibfnamefont {K.}~\bibnamefont
			{Maki}},\ }\bibfield  {title} {\bibinfo {title} {Thermal conductivity of pure
			type-{II} superconductors in high magnetic fields},\ }\href
	{https://doi.org/10.1103/PhysRev.158.397} {\bibfield  {journal} {\bibinfo
			{journal} {Phys. Rev.}\ }\textbf {\bibinfo {volume} {158}},\ \bibinfo {pages}
		{397} (\bibinfo {year} {1967})}\BibitemShut {NoStop}%
	\bibitem [{\citenamefont {Sologubenko}\ \emph {et~al.}(2002)\citenamefont
		{Sologubenko}, \citenamefont {Jun}, \citenamefont {Kazakov}, \citenamefont
		{Karpinski},\ and\ \citenamefont {Ott}}]{sologubenko2002}%
	\BibitemOpen
	\bibfield  {author} {\bibinfo {author} {\bibfnamefont {A.}~\bibnamefont
			{Sologubenko}}, \bibinfo {author} {\bibfnamefont {J.}~\bibnamefont {Jun}},
		\bibinfo {author} {\bibfnamefont {S.}~\bibnamefont {Kazakov}}, \bibinfo
		{author} {\bibfnamefont {J.}~\bibnamefont {Karpinski}},\ and\ \bibinfo
		{author} {\bibfnamefont {H.}~\bibnamefont {Ott}},\ }\bibfield  {title}
	{\bibinfo {title} {Thermal conductivity of single-crystalline
			$\mathrm{MgB}_2$},\ }\href {https://doi.org/10.1103/PhysRevB.66.014504}
	{\bibfield  {journal} {\bibinfo  {journal} {Phys. Rev. B}\ }\textbf {\bibinfo
			{volume} {66}},\ \bibinfo {pages} {014504} (\bibinfo {year}
		{2002})}\BibitemShut {NoStop}%
\end{thebibliography}
\end{document}